\documentclass[conference]{IEEEtran}
\IEEEoverridecommandlockouts
\usepackage{cite}
\usepackage{amsmath,amssymb,amsfonts}
\usepackage{algorithmic}
\usepackage{adjustbox}
\usepackage{graphicx}
\usepackage{caption}
\usepackage{textcomp}
\usepackage{multirow}
\usepackage{comment}
\usepackage{xcolor}
\usepackage[final]{changes}
\usepackage{float}

\usepackage{hyperref}

\makeatletter
\newcommand{\printfnsymbol}[1]{%
  \textsuperscript{\@fnsymbol{#1}}%
}
\makeatother

\def\BibTeX{{\rm B\kern-.05em{\sc i\kern-.025em b}\kern-.08em
    T\kern-.1667em\lower.7ex\hbox{E}\kern-.125emX}}
\begin{document}

\title{A Framework for Crop Price Forecasting in Emerging Economies by Analyzing the Quality of Time-series Data\\
}
\author{\IEEEauthorblockN{Ayush Jain\printfnsymbol{1}\thanks{\printfnsymbol{1}The first two authors contributed equally to this paper.}}
\IEEEauthorblockA{\textit{IBM Research - India}\\
ayjai121@in.ibm.com}
\and
\IEEEauthorblockN{Smit Marvaniya\printfnsymbol{1}}
\IEEEauthorblockA{\textit{IBM Research - India}\\
smarvani@in.ibm.com}
\and
\IEEEauthorblockN{Shantanu Godbole}
\IEEEauthorblockA{\textit{IBM Research - India}\\
shantanugodbole@in.ibm.com}
\and
\IEEEauthorblockN{Vitobha Munigala}
\IEEEauthorblockA{\textit{IBM Research - India}\\
vmunig10@in.ibm.com}
% \and
% \and
}
%\begin{comment}
% \author{\IEEEauthorblockN{Anonymous}
% \IEEEauthorblockA{\textit{Anonymous} \\
% \textit{Anonymous}\\
% Anonymous}
%\and
%\and
%\and
%\and
%\and
%\IEEEauthorblockA{\textit{dept. name of %organization (of Aff.)} \\
%\end{comment}

\maketitle

\begin{abstract}
% crop price forecasting estimates future prices of the crops based on their historical prices. 
Accuracy of crop price forecasting techniques is important because it enables the supply chain planners and government bodies to take appropriate actions by estimating market factors such as demand and supply. In emerging economies such as India, the crop prices at marketplaces are manually entered every day, which can be prone to human-induced errors like the entry of incorrect data or entry of no data for many days. In addition to such human prone errors, the fluctuations in the prices itself make the creation of stable and robust forecasting solution a challenging task. Considering such complexities in crop price forecasting, in this paper, we present techniques to build robust crop price prediction models considering various features such as (i) historical price and market arrival quantity of crops, (ii) historical weather data that influence crop production and transportation, (iii) data quality-related features obtained by performing statistical analysis. We additionally propose a framework for context-based model selection and retraining considering factors such as model stability, data quality metrics, and trend analysis of crop prices. To show the efficacy of the proposed approach, we show experimental results on two crops - Tomato and Maize for 14 marketplaces in India and demonstrate that the proposed approach not only improves accuracy metrics significantly when compared against the standard forecasting techniques but also provides robust models.
\end{abstract}

\begin{IEEEkeywords}
Time-series Data Analysis, Crop Price Prediction, Data Quality of Time-series, Context-based Model Selection
\end{IEEEkeywords}

\section{Introduction}\label{Introduction}
India is an agriculture-based country where 54.6\% of the total workforce is engaged in agricultural and allied sector activities, accounting for 17.1\% of the country’s Gross Value Added (GVA)\footnote{Annual Report 2018-19, Ministry of Agriculture and Farmers Welfare, Government of India}. Hence, it becomes important for the government bodies associated with agriculture to estimate market factors and take suitable actions to benefit the farmers. Therefore, having a robust automated solution, especially in developing countries such as India, not only aids the government in taking decisions in a timely manner but also helps in positively affecting the large demographics.  The price of crops is one such market factor that requires the attention of the government. Accurate crop price forecasting can be useful for the government to take proactive steps and decide various policy measures such as adjusting MSP (Minimum Support Price) so that farmers get a decent price for their produce, restricting the export price by imposing an MEP (Minimum Export Price), so that exporters are forced to sell locally, thus bringing down the crop prices. At the same time, it will also be useful for the farmer for making better decisions like when to sell their produce or when to harvest the crop.

The crop prices are affected due to several factors such as the area under cultivation for a particular crop, supply projection, government policies, consumer demands, supply chain aspects of producers for agriculture-based products, etc. 
%Commodity arrival volume to the marketplace market is another extremely important factor which is captured at the marketplace level throughout the country in India. %Harvested crops are brought to registered marketplaces called marketplaces every day, from where they are sold. The arrival prices of the crops are registered at the marketplace level and closely monitored by government agencies. 
Additionally, weather conditions also play an important factor since the majority of agricultural production in India is rain-fed. Therefore, the study of fluctuations in agricultural crop prices is interesting as well as an important problem to solve from the government's perspective.

Apart from the above-stated reasons, agricultural crop price forecasting is quite challenging due to many factors such as data quality issues, unreliability in future weather predictions, high fluctuation present in the historical crop price, crop price variations across neighboring marketplaces, etc. Moreover, the manually recorded data is prone to human-induced errors such as no data or wrong data entered for a certain day. Considering ML/DL based models, with a new price data arrival every day, updating the models might cause stability issues because of quality issues associated with the crop price data.
% In this paper we present the architecture for robust crop price prediction by analyzing the historical crop price and arrival quantity data, historical weather data  and data quality related features obtained by performing statistical analysis. We propose a framework for context-based model retrieval and retraining based on crop price trend and show experimental results for Tomatoand maize crop price forecasting in some marketplaces. Our results show that our proposed approach improves the accuracy numbers when compared against some standard techniques and provides more robust crop price forecasting.

In this paper, we address the set of challenges present in deploying a real-world crop price prediction solution by evaluating its robustness and reliability over a continuous time-frame. 
%and capture our approach in Figure \ref{fig:approach}. 
We summarize our contributions as below: 
%and show overview of our approach in Fig. \ref{fig:approach}.
% To deal with the anomalies present in the data and variations present in the weather conditions, we propose a multivariate setting which uses the weather data, Agmarknet data and data quality related factors. The fluctuation in the crop prices is addressed by context-based model selection framework which also addresses the issues related to model stability.  Figure \ref{fig:approach} shows the overall steps of our proposed approach. 
%In this paper, we present a deployed crop price prediction framework that is evaluated based on reliability and robustness over a continuous time-frame. 
% Our contributions are as follows:  
\begin{itemize}
    \item We present the architecture of end-to-end pipeline for robust crop price prediction by analyzing historical marketplace data, weather data and data quality-related features.  
    % \item We propose a framework for enabling context-based model selection strategies under different conditions such as identifying context-based on data quality metrics, model stability, and historical crop price trend. 
    % \item We present results with different types of time-series regressor models and study the impact of building models for each market versus each crop to study the reliability and robustness of the model across time.
    \item We propose a framework for enabling context-based model selection strategies under different conditions such as identifying context based on data quality metrics, model stability, and trend of historical crop prices. 
    %\item We present results with different types of time-series regressor models and study the impact of building models for each market versus each crop to study the reliability and robustness of the model across time.  
    \item We experiment with various regression models and show the results for two crops namely, Tomato and Maize for 14 marketplaces in India. We additionally report interesting findings to illustrate the benefits in modeling trend specific models built on marketplace level as well as crop level for a robust crop price forecast.  
    %\item We propose a condition continuous retraining mechanism to deal with data quality issues to improve the robustness of the time-series regressor techniques. 
\end{itemize}

% \begin{figure}[htbp]
% \centerline{\includegraphics{figures/overall_framework_commodity_price_prediction.png}}
% 	\caption{Overall steps of our approach for crop price prediction framework. \label{fig:approach}}
% \end{figure}

\section{Related Work} \label{Related work}
Time-series forecasting has been an active research area and has been studied under various scenarios such as stock price prediction \cite{Bao2004ForecastingSP}, energy load forecasting \cite{Amarasinghe2017DeepNN}, traffic forecasting \cite{Li2017DiffusionCR}, crop yield prediction \cite{Shahhosseini2020ForecastingCY}, etc.  To capture time-series specific properties, various categories of models were proposed such as (i) Univariate models, which can only model the endogenous variables, like MA, ARMA, ARIMA and its variants such as SARIMA (ii) Multivariate models which can model exogenous variables along with the endogenous variables like VAR models along with its variants such as elliptical \cite{Qiu2015RobustEO}, structured VAR model \cite{Melnyk2016EstimatingSV}, ARIMAX, SARIMAX, etc. Along with these time-series specific models, many classical machine learning models found their way into the problem of time-series forecasting such as support vector regression \cite{KIM2003307}, LASSO \cite{LI2014996}, gaussian processes \cite{rsta2011} and even recent deep learning techniques such as LSTMs \cite{SiamiNamini2018ForecastingEA}. Apart from using single models, various ensemble models \cite{OliveiraT14,CerqueiraTPS17,ShenBAW13} were designed and tested to improve the accuracy of time-series forecasting problems.\\
\hspace*{4mm} Crop price prediction is one such instantiation of time-series forecasting problem which is being actively pursued by the research community in the recent past. For example, \cite{bame_2008} use exponential smoothing, ARIMA and spectral methods for predicting crop prices. \cite{Yercan2012} exploit the seasonal properties of the prices and recommend seasonal models like SARIMA for Tomato price prediction. \cite{Gloria2013} proposes spline-based interpolation techniques for Tomato price prediction.  \cite{Nasira2012VegetablePP} proposed usage of Back Propagating Neural Networks (BPNN) for forecasting vegetable prices. \cite{Hemageetha2013RadialBF} modeled the problem of Tomato price forecasting using the radial basis function neural networks. \cite{Ouyang2019} proposed the usage of Long- and Short-Term Time-series Network (LSTNet) for modeling the prices of agricultural crops. In addition to these simple models, various ensemble models were also proposed for efficiently forecasting techniques in agriculture-related problems. \cite{Shahhosseini2020ForecastingCY} used stacking based ensemble models for predicting crop yield. \cite{Xiong2018SeasonalFO,taylor2018forecasting} decompose time-series data into a seasonal, trend, and remainder components and proposed individual model to forecast the trend, remainder and seasonal components and finally add them to get the final forecast value. However, most of the prior works focus on continuous retraining of the model as opposed to contextual model retraining. 

Further, apart from concentrating on building complex models, researchers proposed various additional features that can help the price forecasting problem. \cite{chakraborty2016predicting} analyzed new events along with the Agmarknet data to predict food prices and reported a significant improvement over the standard ARIMA model. \cite{Ma2018AnIP} used collaborative filtering for imputing missing data values using data from neighboring marketplaces. \cite{madaan2019price} performed price forecasting and anomaly detection in the crop prices by training a classifier model on the dataset of news articles covering hoarding related incidents. \cite{Zhang2020ForecastingAC} proposed various features such as complexity, linearity, stationarity, periodicity in addition to a model based on horizon features to model agricultural prices. %\cite{soyabean_2019} uses features such as historical yield, production, exchange rate, historical rainfall, etc for price prediction of soybean and onion.  
A pilot study\footnote{Price Predictions using Machine Learning (AI) for Soyabean and Onion, Atal Bihari Vajpayee Institute of Good Governance and Policy Analysis, 2019} was performed on soybean and onion for price prediction that uses features such as historical yield, production, exchange rate, historical rainfall, etc.  \cite{Kantanantha2010} uses environmental factors like rainfall and temperature for price prediction of corn and soybean. However, these prior works do not consider features related to data quality aspects.

In our work, we explore the time-series forecasting models proposed in the literature as discussed above in the context of crop price prediction. Specifically, we analyze the impact of data quality of time-series data and properties of the time-series data such as trend on the downstream models and suggest a framework for continuous retraining of models upon arrival of new data and retrieval of models for the best forecast. Additionally, we also study the impact of building models for each market versus each crop and report our findings in the form of quantitative and qualitative results.

\begin{figure*}[t]
  \centering
  % % \vspace{-4mm}
  \begin{minipage}[b]{0.49\textwidth}
    \includegraphics[width=\textwidth]{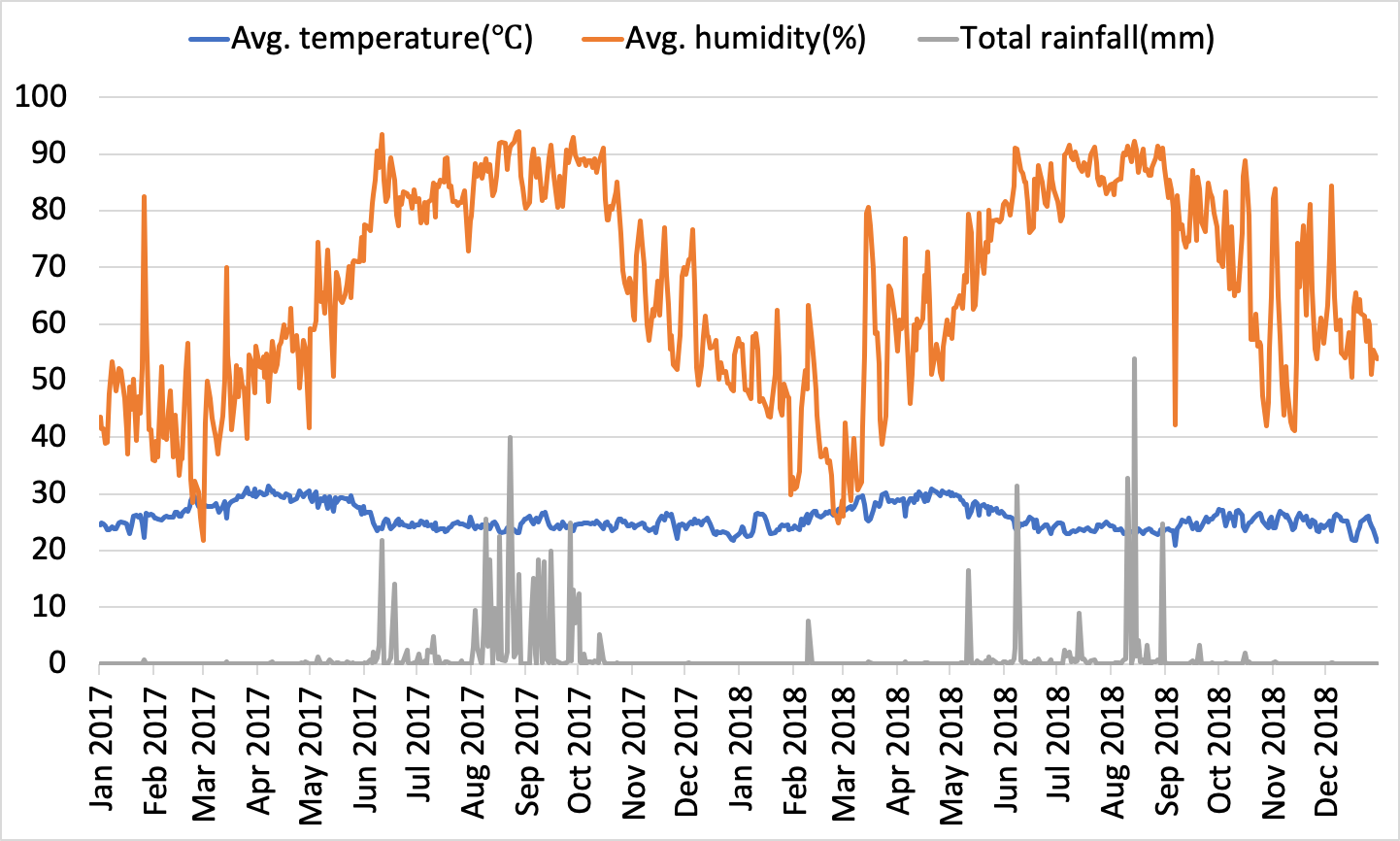}
    % % \vspace{-6mm}
    \caption*{(a) Weather parameters (Davangere)}
  \end{minipage}
  \hfill
  \begin{minipage}[b]{0.49\textwidth}
    \includegraphics[width=\textwidth]{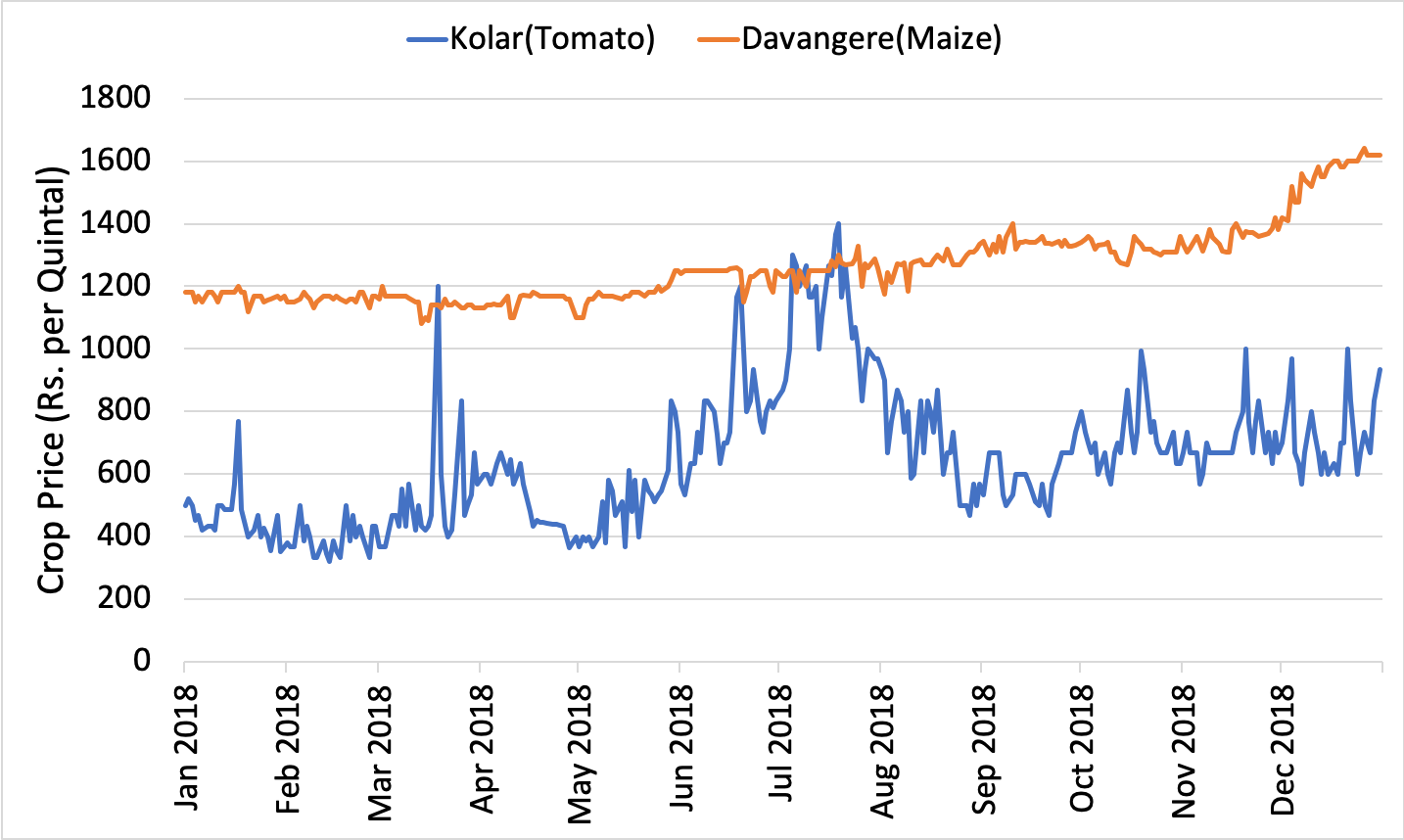}
    % % \vspace{-6mm}
    \caption*{(b) Crop prices}
  \end{minipage}
  % % \vspace{-2mm}
  \caption{Variations present in weather condition shown in (a) and crop price variations shown in (b). }
  % % \vspace{-8mm}
  \label{fig:weather_price}
\end{figure*}

\begin{figure*}
  \centering
  \begin{minipage}[b]{0.49\textwidth}
    \includegraphics[width=\textwidth]{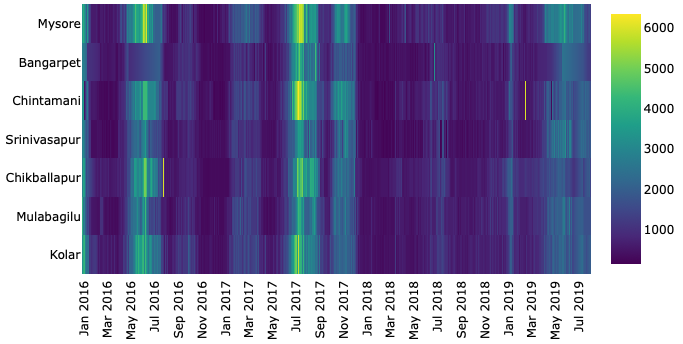}
     \vspace{-6mm}
    % \captionsetup{labelformat=empty}
    \caption*{(a) Tomato}
  \end{minipage}
  \hfill
  \begin{minipage}[b]{0.49\textwidth}
    \includegraphics[width=\textwidth]{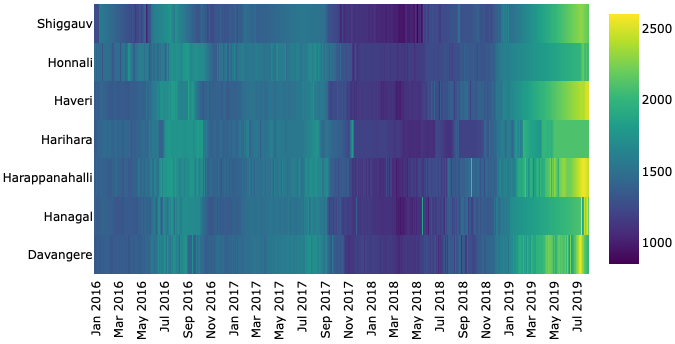}
     \vspace{-6mm}
    %  \captionsetup{labelformat=empty}
    \caption*{(b) Maize}
  \end{minipage}
   % \vspace{-3mm}
  \caption{Historical crop price variations for Tomato and Maize \added{Prices in Rs. per Quintal}.}
  % \vspace{-5mm}
  \label{heat_map}
\end{figure*}

\section{Dataset} \label{Dataset}
In this section, we briefly discuss various factors we use for modeling crop prices and provide necessary evidence to showcase their importance. In our analysis, we choose different regions from Karnataka state in India for analyzing the prices of Tomato and Maize.

\subsection{Weather Data}\label{weather}
Crop prices are affected due to many factors such as (i) bad weather like heavy rainfall may impact the routes \added{used for transporting crops to the marketplaces} (ii) temperature and humidity affect the shelf life of crops (iii) quantity of crops brought to the marketplaces, as a surplus supply generally brings the prices down and vice versa, which is in turn affected by the weather conditions. Therefore it becomes important to consider weather-related information while predicting prices of crops.
We analyze weather parameters such as average humidity, total rainfall and average temperature on a daily basis. We use The Weather Channel APIs to access the weather data for the geo-locations of all the marketplaces. Fig. \ref{fig:weather_price}(a) shows the variations in the weather parameters for Davangere district in Karnataka state for the years 2017 and 2018.

\subsection{Agmarknet Data}
Agmarknet\footnote{https://agmarknet.gov.in/}(Agricultural Marketing Information Network) website is maintained by the Government of India that maintains crop related information such as minimum, maximum, and the modal price per quintal at marketplace level. \added{The Agmarknet website also captures the marketplace arrival volume of each crop which is measured in quintal for every marketplace. Agmarknet website is updated every day except Sundays since marketplaces are closed and there is no arrival of crops on Sundays. Sometimes, even for the weekdays, data is not updated on the website due to some issues resulting in days with missing values.}

%\added{ along with the marketplace arrival volume \footnote{The marketplace arrival volumn of the crop is measured in quintal at daily basis.} of a crop.} 
%This website gets updated daily and captures the crop arrival volume information as well.
%At the marketplace level, the arrival volume of a crop along with the  minimum,  maximum  and  the  modal  price  per  quintal is registered.  The information is updated every day with  new arrivals. This data is available on the Agmarknet(Agricultural Marketing Information Network) website run by the government of India.
Fig.  \ref{fig:weather_price}(b) shows the modal price variations for Tomato and Maize from Kolar and Davangere marketplaces in Karnataka. It can be observed that the variations in prices are much more pronounced in Tomato than in Maize.

Fig.  \ref{heat_map} shows the heat maps\footnote{The missing values have been imputed as provided in \ref{stat_ana}} \added{which visualize the variation in modal price per Quintal of Tomato and Maize} for  marketplaces \added{located in different districts in Karnataka.}
% \footnote{We would henceforth refer to the marketplaces by the name of the districts in which they are located}
From the heat maps, we can infer that there is some seasonality in the prices that can be seen from the repeating patterns in any row over a period of 3 and a half years. Also, it can be seen that some of the marketplaces appear to have correlation amongst them.
\begin{comment}
\added{
\subsection{Data Quality Information}
As mentioned earlier, since the data is captured manually at each marketplace, it suffers from data quality issues such as missing values or outlier values on certain days. Therefore the Agmarknet data is analyzed to assess its quality and perform suitable processing before we proceed to the task of price prediction. We consider this data quality analysis as an additional source of information to construct new features which can then be provided to the forecasting models.
}
\end{comment}
\begin{figure*}
    \centering
    % % \vspace{-2mm}
	\includegraphics[width=\linewidth]{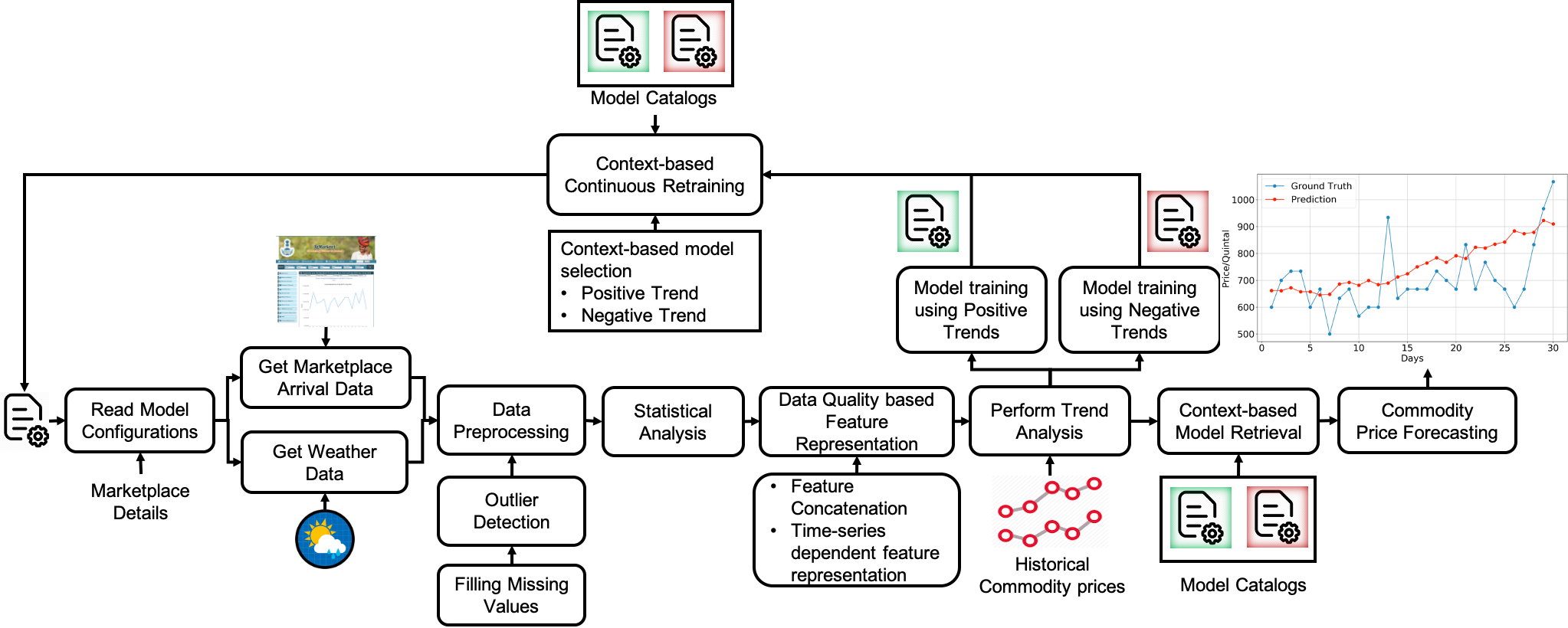}
	% % \vspace{-7mm}
	\caption{Overall steps of our approach for crop price prediction framework using a trend-based model selection strategy. \label{fig:approach}}
	% % \vspace{-4mm}
\end{figure*}
% \begin{figure*}[!htb]
%   \centering
%   \begin{minipage}[b]{0.45\textwidth}
%     \includegraphics[width=\textwidth]{figures/miss_out_tomato.png}
%     %vspace{-7mm}
%     \caption*{(a) Tomato}
%   \end{minipage}
%   \hfill
%   \begin{minipage}[b]{0.45\textwidth}
%     \includegraphics[width=\textwidth]{figures/miss_out_maize.png}
%     %vspace{-7mm}
%     \caption*{(b) Maize}
%   \end{minipage}
%   %vspace{-3mm}
%   \caption{Percentage of missing and outlier values for Tomato and Maize}
%   \label{miss_out}
% \end{figure*}
\begin{figure}[!htb]
  \centering
    \includegraphics[width=0.5\textwidth]{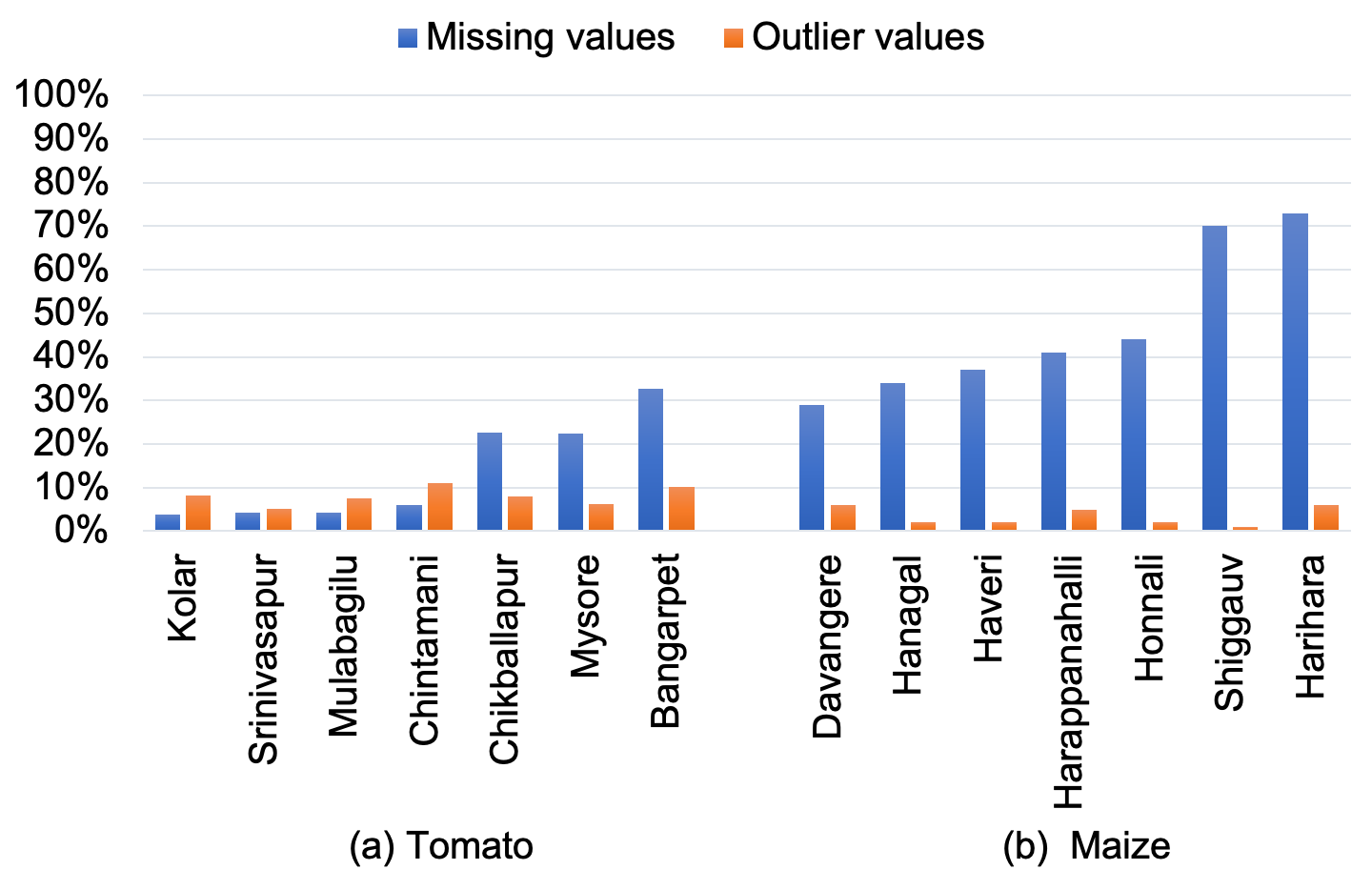}
    %vspace{-7mm}
  \vspace{-6mm}
  \caption{Percentage of missing and outlier values for Tomato and Maize.}
  \label{miss_out}
\end{figure}
\section{\added{Proposed Method}}
\begin{comment}
\added{In this section, we describe how the data quality and statistical analysis was performed on the Agmarknet data, how the features are represented to the model and finally how the context based model selection strategies were designed to address the shortcomings identified in data quality analysis.}
\end{comment}
\added{Given historical crop prices along with weather data and other Agmarknet data, we first perform data preprocessing steps to deal with missing values and outliers. We then perform statistical analysis on the crop price data to obtain insights into the data quality along with trend and variations present in the data. We identify a set of additional features with the help of the insights from the statistical analysis. Two different feature representation techniques are introduced to capture data quality aspects along with Agmarknet data and weather data.  The data quality based feature representation and the context-based model selection strategies have been designed to deal with data quality issues in time-series forecasting. 
%and statistical analysis to extract data quality based feature representations using two different techniques followed by a set of different context-based model selection strategies. %by performing statistical tests to measure the stationarity of time-series data.  
Fig. \ref{fig:approach} shows a high-level pipeline of our proposed framework using trend-based model selection strategy.}

\begin{comment}
\deleted{We performed rigorous statistical analysis on the Agmarknet data with two objectives. Firstly, as explained in the previous sections, the Agmarknet data is entered manually into the system and suffers from quality issues such as missing values on certain days, presence of outlier data among others. Secondly, some aspects of time-series data, such as stationarity, cannot be ascertained by simply plotting the data. In this section, we discuss the pre-processing techniques and statistical tests along with the insights that helped us in constructing features in addition to the weather and Agmarknet data discussed in the previous section. Finally, we discuss the feature representations that were used for price forecasting.}
\end{comment}

\subsection{\added{Data Preprocessing and }Statistical Analysis} \label{stat_ana}
\added{In the data quality analysis, we primarily focus on the identification of missing values and outliers present in the data. All the days (except Sundays) where there was no entry were considered as missing values 
% \added{\footnote{One can use additional information of nearby marketplaces for filling the missing values for crop prices similar to \cite{Ma2018AnIP}.}}
in the Agmarknet data and filled using spline based technique \cite{junninen2004methods}. Outliers in the data were detected using IQR (Interquartile range) method \cite{barbato2011features} with threshold value as 1,  and the outlier values were not modified. We attribute the presence of outliers to human errors while entering the data in a digital system.}  Fig. \ref{miss_out}(a) and \ref{miss_out}(b) show the fraction of days with missing values and outliers for the considered marketplaces of Tomato and Maize respectively.
% \added{We ignore marketplaces for Tomato such as Malur and Gowribidanoor that have a high number of missing values (above 70\%) from our analysis since it will be difficult to capture the price variations efficiently. Similarly, we ignore marketplaces such as Savanur and Channagiri for Maize that a have high number of missing values (above 80\%).}

\added{In the statistical analysis,} we perform \textit{ADF},  \textit{KPSS} tests \cite{schlitzer1995testing} and compare the test statistics with the 5 percent critical value. Based on these two tests, we infer if the data is non-stationary, strict-stationary, trend-stationary or difference-stationary. Time-series for tomato price is found to be strict stationary while that of maize is found to be non-stationary. \added{Further, the Agmarknet data is decomposed into trend, seasonal and residual components using additive seasonal decomposition techniques.}
\begin{comment}
We found the trend and seasonality strength by using equations \ref{tre_stre} and \ref{sea_stre}.  Here $R_t$ and $T_t$ refer to the residual and trend components respectively. 
% \begin{equation}\label{tre_stre}
%     F_T = \max\left(0, 1 - \frac{\textit{Var}(R_t)}{\textit{Var}(T_t+R_t)}\right)
% \end{equation}
% \begin{equation}\label{sea_stre}
%     F_S = \max\left(0, 1 - \frac{\textit{Var}(R_t)}{\textit{Var}(S_{t}+R_t)}\right)
% \end{equation}
\newline
\noindent\begin{minipage}{.5\linewidth}
\begin{equation}\label{tre_stre}
F_T = \max\left(0, 1 - \frac{\textit{Var}(R_t)}{\textit{Var}(T_t+R_t)}\right)
\end{equation}
\end{minipage}%
\begin{minipage}{.5\linewidth}
\begin{equation}\label{sea_stre}
 F_S = \max\left(0, 1 - \frac{\textit{Var}(R_t)}{\textit{Var}(S_{t}+R_t)}\right)
\end{equation}
\end{minipage}
\end{comment}

We further analyze \added{the statistical properties of the} residual component or noise. The mean is quite close to zero for all marketplaces which is expected. The average standard deviation \added{of the residual component for} Tomato is 202.43 which is much higher than that of Maize, which is found to be 34.39, \added{indicating that there is} a higher fluctuation in Tomato prices than in Maize. 
\subsection{\added{Data Quality based Features}} \label{insight}
\added{Based on the insights from the data quality and statistical analysis, the following list of features is proposed to capture the data quality issues and variations present in the Agmarknet time-series data.}

%Based on the insights, we obtained during data pre-processing and statistical analysis, we add the following features based on data quality and variations in crop price time-series data.}

\noindent \textbf{Missing value flag (M)}:
\added{From the missing value analysis, a \textit{missing value flag} is added as a feature, encoded as a binary value of 1 for days with missing values and 0 for others.}  \newline %This feature was assigned a 1 for all the days with imputed values and 0 for the other days.
\textbf{Outlier flag (O)}: \added{From outlier analysis, an \textit{outlier flag} is added as a feature. The flag takes value 1 for all the days which were identified as outliers and 0 for the other days.} \newline
% \textbf{Fourier transform based features (FT)}:
% FFT (Fast Fourier Transform) is performed on the Agmarknet data, thereby transforming it into the frequency domain. Further, Inverse FFT was performed after rejecting higher frequencies\cite{fumi2013fourier} and the resulting time-series were used as features. In our experiments, analysis was done for the first 3, 6, 9, and 100 frequencies and it was observed that transforms with more components were closer to the real data. These additional features help in removing the noise component from the data and discover the underlying patterns.\newline
\textbf{Fourier transform based features (FT)}:
FFT (Fast Fourier Transform) is performed on Agmarknet data, thereby transforming it into the frequency domain. Then, higher frequency components are removed from the FFT, following which Inverse FFT is performed to obtain time-series which is used as features \cite{fumi2013fourier}. In our experiments, we retain first 3, 6, 9, and 100 frequencies to obtain 4 time-series inverse transforms, with the transform with more components closer to the real data. These additional features help in removing the noise component from the data and discover the underlying patterns.\newline
%Figure \ref{fft_comp} shows actual data along with the inverse fourier components. As expected the transforms with more components are closer to the real data. \newline
% \begin{figure}[!tbp]
%   \centering
%   \begin{minipage}[b]{0.49\textwidth}
%     \includegraphics[width=\textwidth]{figures/fft_kolar.png}
%     % \vspace{-8mm}
%     \caption*{(a)}
%   \end{minipage}
%   \hfill
%   \begin{minipage}[b]{0.49\textwidth}
%     \includegraphics[width=\textwidth]{figures/fft_davangere.png}
%     % \vspace{-8mm}
%     \caption*{(b)}
%   \end{minipage}
%   % \vspace{-4mm}
%   \caption{Fourier transform components for Kolar(a) and Davangere(b)}
%   \label{fft_comp}
% \end{figure}
\textbf{Time-related features (T)}:
Information such as day and month of crop arrival are considered as additional features inorder to capture the temporal seasonality in the prices. These features are encoded in numeric form.\newline
% \subsubsection{History parameters}
% Observing the high variance in the data we see intuitively that there the marketplace price of a particular day should be more dependent on recent past prices. This aspect was also explored in our experimentation.
\textbf{Statistical Indicators (SI)}:
Statistical indicators such as moving averages, moving standard deviation of the price time-series are added as a part of the feature set. These indicators are useful in capturing recent variations in time-series for long-term forecasting and are usually used in problems related to stock price predictions. \added{In our experiments, we use the following indicators: 3 and 7 days Moving Averages, Exponential Moving Averages with Center of Mass 0.25 and 0.5, 20 days Moving Standard Deviation and Moving Average Convergence Divergence (MACD), as used in \cite{thomas2019time}}.
%  \added{In our experiments, following values are used for the indicators: 7 and 21 days moving averages, 12 and 26 days exponential moving averages, 20 days moving standard deviation and exponential weighted mean as mentioned in \cite{agrawal2019stock}}

\begin{comment}
\begin{table}[t]
\tiny
\caption{List of statistical indicators used along with their definitions}
\label{ind_table}
\centering
\begin{tabular}{|l|l|}
\hline
 MA7 & 7 days moving average\\
 \hline
 MA21  & 21 days moving average  \\
\hline
12ema & 12 days exponential moving average\\
\hline
26ema & 26 days exponential moving average\\
\hline
MACD & Moving average convergence and divergence (12ema - 26ema)\\ \hline
20std & 20 days moving standard deviation \\ \hline
Upper band & MA21+ 2*20std\\ \hline
Lower band & MA21- 2*20std \\ \hline
Momentum & Price-1  \\ \hline
Log momentum & Logarithm of momentum to base 10 \\ \hline
EMA & Exponential weighted mean \\ \hline
\end{tabular}
\end{table}
\end{comment}
\subsection{Feature Representation \added{}} \label{feat_rep}
\begin{comment}
For training we have parameters called \textit{n{\_}steps{\_}in} and \textit{n{\_}steps{\_}out} shown as $p$ and $q$ in figure \ref{data_pr} respectively. \textit{n{\_}steps{\_}in} number of days are used to predict the prices for the next \textit{n{\_}steps{\_}out} days. The feature representation is prepared as shown in figure \ref{data_pr}. Suppose we have \textit{m} features represented as $f_1$, $f_2$ to $f_m$ (like arrival quantity, humidity etc.) along with arrival price $c$ for n days. Let us number the days from ${D^1}$ to ${D^n}$. Hence for each day $D^i$, we have feature set as ($x^i_1, x^i_2, ..., x^i_m, c^i $). This feature set is represented as $X\_D^i$
% This data is split between training and test data. ${D_1}$ to ${D_k}$ are taken in the training set and ${D_{k+1}}$ to ${D_n}$ are taken in test set.
Now we take days from ${D^1}$ to ${D^p}$ and concatenate their features set $X\_D^1$, $X\_D^2$ to $X\_D^p$. This creates a data with \textit{m}*\textit{n{\_}steps{\_}in} features and becomes ${X^1}$. For creating label $Y^1$, the crop prices from day ${D^{p+1}}$ to ${D^{p+q}}$ are concatenated. $X^1$ and $Y^1$ constitute one training data point. This method is followed for the entire for the entire dataset ans is shown in figure \ref{data_pr}.
\end{comment}
\begin{comment}
\begin{figure}
	\centering
	\includegraphics[width=0.7\textwidth]{figures/feat_rep.png}
	%\r4trret4{-0.5mm}
	\caption{Figure showing preparation of training and test data}
	\label{data_pr}
	%\r4trret4{-4.5mm}
\end{figure}
\end{comment}

Typically, historical crop prices are used to obtain the feature representation. In addition to the Agmarknet data ($AG$), we utilize additional information such as weather data ($WD$), and data quality-related features as discussed above ($DQ$) for constructing the final feature representation. All the individual features can be combined into single representation using the two definitions of $\mathcal{F}$ as described below:
\begin{enumerate}
	\item \textbf{Feature Concatenation\label{feat_concat}:} \added{To construct a feature vector for $i^{th}$ day,} we concatenate the historical $k$ days of attributes and use it to train a regressor model for crop price forecasting. Let $f$ be the operator that creates the feature representation $\mathcal{F}(d_{i})$ for $i^{th}$ day. For $f_{AG}(d_{i, i-k})$ concatenate arrival quantity and crop price from $i^{th}$ day to $(i-k)^{th}$ day. Similarly, we compute the feature representation for weather data ($f_{WD}(d_{i, i-k})$), and feature representation for data quality ($f_{DQ}(d_{i, i-k})$) for $i^{th}$ day.  The concatenated features for $i^{th}$ day is computed as:
\begin{equation}
\mathcal{F}(d_i) = [f_{AG}(d_{i, i-k}), f_{WD}(d_{i, i-k}), f_{DQ}(d_{i, i-k})] \label{eq:feature_concate}
\end{equation} 
\added{The feature representation $\mathcal{F}(d)$ is a combination of WD, DQ, and AG based historical features. The final combined feature representation is of dimension size $k\cdot$$N_{WD}$ + $k\cdot$$N_{DQ}$ + $k\cdot$$N_{AG}$ dimensional, where $N_{WD}$, $N_{DQ}$ and $N_{AG}$ are the dimensions of daily weather data, data quality, and Agmarknet data respectively. The values of $N_{WD}$ is 3, $N_{DQ}$ is 13, and $N_{AG}$ is 2.  To show the importance of the constructed feature vector, we use a Multilayer perceptron (MLP) regressor with 1 hidden layer containing 10 neurons with \textit{ReLU} activation, using feature concatenation-based feature representation. To train the model we use lbfgs solver with a constant learning rate of 0.001 and squared error as the loss function. The MLP model takes as input the feature vector $\mathcal{F}(d_{i})$ and makes a forecast for the next 30 days i.e., ($i+1$) to ($i+30$) days.} 
\begin{figure}[tb]
\centering
% % 	% \vspace{-9mm}
% 	\includegraphics[width=0.65\textwidth]{figures/time_series_dependent_feature_representation.png}
	\centerline{\includegraphics[width=\linewidth]{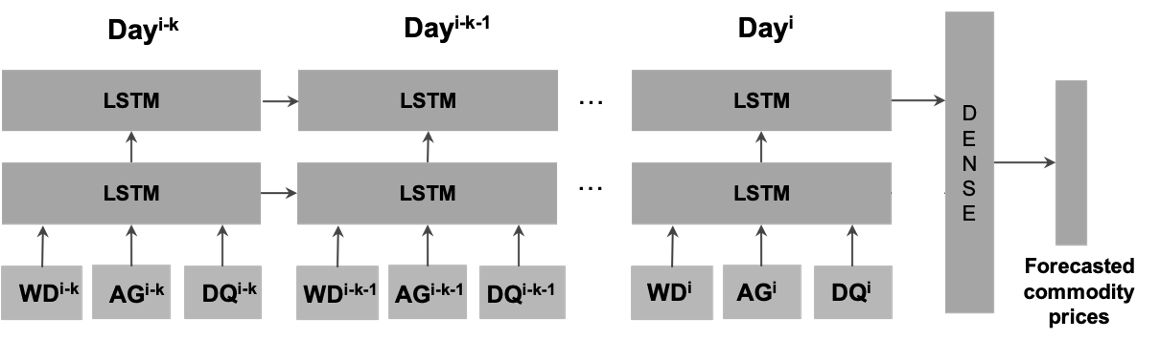}}
	\caption{LSTM-based feature encoder for crop price forecasting}
% 	% \vspace{-9mm}
	\label{lstm_feat_rep}
\end{figure}
	\item \textbf{Time-series dependent feature representation}: 
	To capture the time-series dependency between the different attributes, we use a stacked LSTM based feature encoder to represent the multivariate data \cite{SiamiNamini2018ForecastingEA}. \added{The LSTM model has a stack of two LSTM layers to encode the input features and the output from the last time step of the second LSTM layer is considered as an input to a Dense layer which forecasts for 30 days. The LSTM layer encodes the sequential information from the input that captures the historical weather data, data quality-related features, and Agmarknet data through the recurrent network. Fig. \ref{lstm_feat_rep} shows the high-level steps for extracting time-series dependent feature representation and use it for crop price forecasting.} \added{In our experiments, the two LSTM layers are 100 and 100 units respectively with each layer having \textit{ReLU} activation. To train the network, we use \textit{Adam} optimizer with a learning rate of 0.001 and mean squared error as the loss function.} %We use these parameters to report the experimental results.
\end{enumerate}

%\added{Given the two different ways of feature construction techniques, one can build two different types of regressor models.  In our experiments, we build a Multilayer perceptron (MLP) regressor using feature concatenation-based technique.  Similarly, we build a stacked-LSTM regressor using time-series dependent feature representation. 

\added{Next, we introduce different contexts to retrain and select models considering different factors such as data quality, model stability, and trend analysis. The procedure for model retraining, model management, and version control for different model selection strategies for a given context is also described.}
\subsection{Context-based Model Selection Strategies} \label{context_mod_sel}
\begin{figure*}
	\centering
    % \vspace{-1mm}s
	\includegraphics[width=0.8\textwidth]{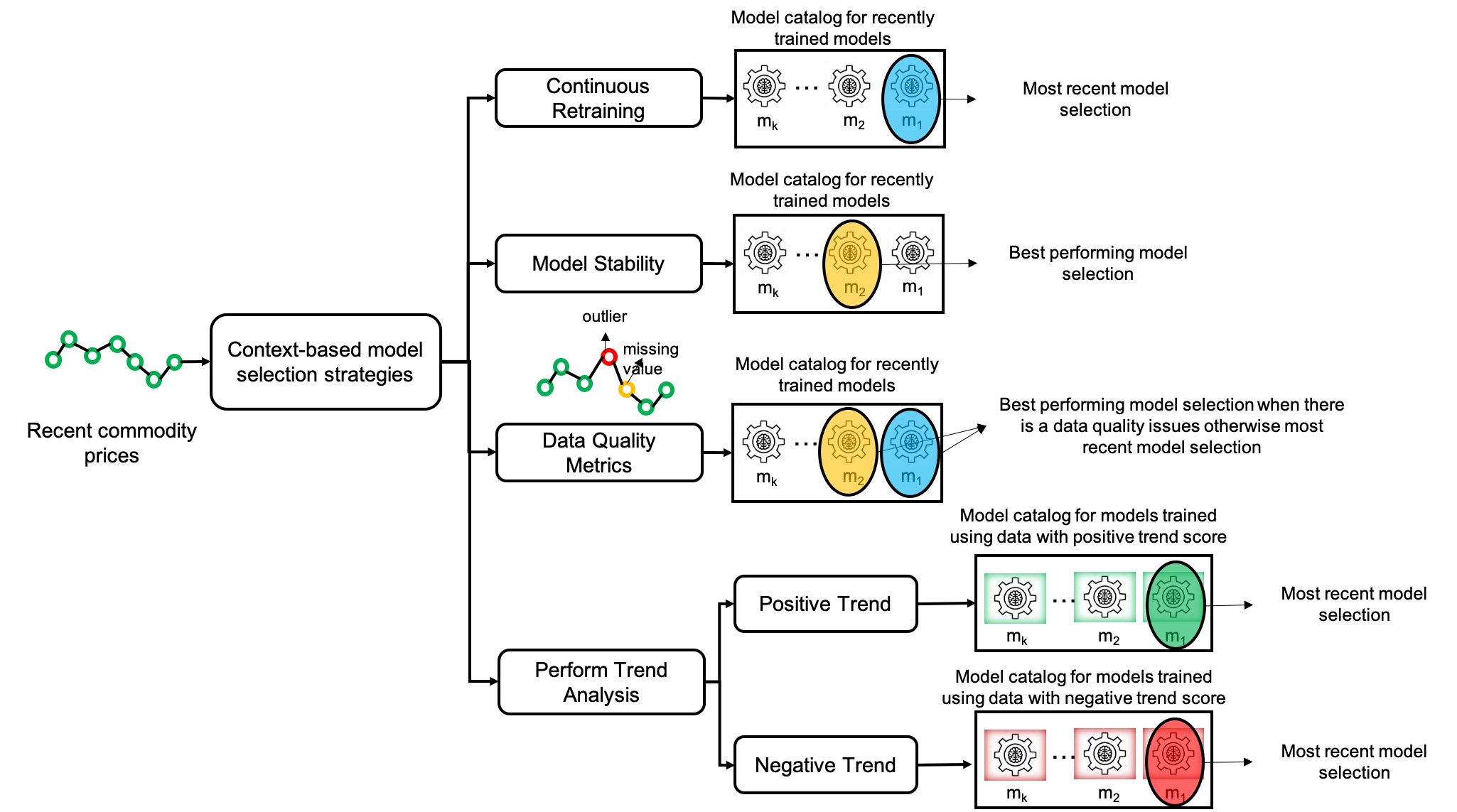}
	% \vspace{-3mm}
	\caption{A set of different context-based model selection strategies for crop price forecasting.}
	\label{context_based_model_selection_stratergies}
% 	\vspace{-2mm}
\end{figure*}

\begin{figure}
	\centering
    % \vspace{-1mm}
	\includegraphics[width=0.5\textwidth]{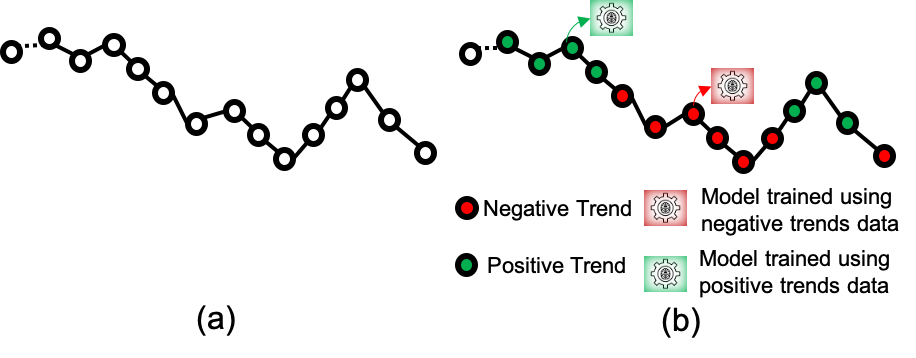}
% 	\vspace{-3mm}
	\caption{\added{An example of model selection using positive and negative trend scores. (a) shows a sample crop price time-series. 
	(b) shows a context-based model selection strategy using trend analysis.} 
	%If the trend score of a data-point is negative (marked with red in (b)) a negative trend model is retrieved, otherwise if the trend score is positive (marked with green in (b)), a positive trend model is retrieved.  
	}
	\label{trend_based_model_selection}
\end{figure}

In real-world time-series data analysis, \added{automated way of model selection and model deployment process} is very important. \added{We propose} a framework for context-based model selection strategies for crop price forecasting.  We also demonstrate the utility of inferring the context by analyzing the variations in recent crop prices along with issues related to model stability and data quality.  Fig. \ref{context_based_model_selection_stratergies} illustrates a set of different model selection strategies for crop price forecasting.
\begin{enumerate}
    \item \textbf{Continuous Retraining:} \label{continuous_retraining}
The model deployment can be treated as a continuous process rather than deploying a model once. One of the most important factors is to enable continuous model enrichment such that the model captures the recent crop price variations while forecasting long-term crop prices.  However, this continuous model enrichment process requires extra processing for model retraining.  Furthermore, continuous retraining may get impacted because of data quality issues if there is no supervision. 
 \item \textbf{Model Stability:} \label{model_stability}
To address the limitation of the continuous retraining process, we automatically analyze the model performance after performing model retraining.  This is achieved by creating a model catalog that stores the set of models along with model performance-related attributes.  The best performing crop price forecasting model is identified as 
%\begin{equation}
    %\centering
	$\arg\max_{m_{i} \in \mathbf{M_{c}}} \phi^{acc}(m_i,D^{val})$
%\end{equation}
where $m_{i}$ represents the a model from the model catalog ($M_{c}$). $\phi^{acc}$ computes the performance metrics on the validation dataset ($D^{val}$). To capture the recent price variations in the model, we put a constraint on the model catalog such that it stores only recently trained models.  
 \item \textbf{Data Quality Metrics:} \label{data_quality_metrics}
The data quality check is very important for model deployment and the model selection process. We analyze the data quality by identifying outliers, consecutive missing values present in the recent time-series, and decide whether to use this data for retraining the model.  This way a continuous condition is enabled to captures the data quality aspects first, and then perform the continuous retraining.  This step could also help in reducing the number of continuous model enrichment steps and thereby reducing the computational cost overall in an automated manner. 
 \item \textbf{Trend Analysis:} \label{trend_analysis}
There is a significant body of work in the space of time-series forecasting using trend, seasonality, seasonal variations, random or irregular movements for modeling time-series data. In this work, we determine trend in crop prices to infer the context. We analyze recent crop prices ($CP_{1,...,k}$) for $k$ days and determine whether there is a positive trend ($T^{pos}$) or a negative trend ($ T^{neg}$) by computing the trend score using:
\begin{equation} \label{eq1}
\small
\begin{split}
T(CP_{1,...,k})= \begin{cases}
 T^{pos} & \sum_{i=1}^{k-1} \frac{CP_{i+1} - CP_{i}} {CP_{i}} > 0 \\ \\
 T^{neg} & \sum_{i=1}^{k-1} \frac{CP_{i+1} - CP_{i}} {CP_{i}} <= 0 \\
\end{cases}
\end{split}
\end{equation}
Two different model catalogs are maintained to capture the positive trend and the negative trend related information. To predict crop prices, a recently trained model is retrieved from the catalog based on the context \added{as shown in Fig. \ref{trend_based_model_selection}} and its effectiveness is discussed in Sec. \ref{exp:context_based_model_selection}. The model catalogs are updated based on the trend analysis of the recent time-series data.  
% \begin{figure}
% \centering
% \begin{subfigure}[b]{0.55\textwidth}
%   \includegraphics[width=1\linewidth]{figures/price_data.png}
%   \caption{}
%   \label{fig:p1} 
% \end{subfigure}
% \begin{subfigure}[b]{0.55\textwidth}
%   \includegraphics[width=1\linewidth]{figures/price_data_trend.png}
%   \caption{}
%   \label{fig:p22}
% \end{subfigure}
% \caption[]{Figure showing the model selection}
% \end{figure}
\begin{comment}
\begin{figure}
  \centering
  \begin{minipage}[b]{0.24\textwidth}
    \includegraphics[width=\textwidth]{figures/price_data.png}
    %vspace{-7mm}
    \caption*{(a)}
  \end{minipage}
  \hfill
  \begin{minipage}[b]{0.24\textwidth}
    \includegraphics[width=\textwidth]{figures/price_data_trend.png}
    %vspace{-7mm}
    \caption*{(b)}
  \end{minipage}
  %vspace{-3mm}
  \caption{Figure showing the selection of positive and negative trend models. (a) shows a sample crop price time-series. If the trend score of a data-point is negative (marked with red in (b)) a negative trend model is retrieved, otherwise if the trend score is positive (marked with green in (b)), a positive trend model is retrieved.  }
  \label{miss_out}
\end{figure}
\end{comment}
\end{enumerate}
\added{ In the next section, we evaluate and compare the different feature representations using continuous retraining strategy. Furthermore, we also show the importance of different context-based model selection strategies such as Model Stability, Data Quality Metrics, and Trend Analysis.}        

%In the next section, we show the experimental results for the price forecasting. We compare the univariate feature representation (using only the historical crop prices) with multivariate feature representation (using AG+WQ, AG+WQ+DQ}. We also compare new strategies B,C and D with the standard strategy A.
\section{Experiments}
We evaluate our crop price forecasting algorithm for \added{7} marketplaces of Tomato and \added{7} of Maize using the Agmarknet dataset. Table \ref{tab:dataset_distribution} shows the train-test split of the datasets for Tomato and Maize. To evaluate the robustness of the crop price forecasting algorithms we first build the model using the historical data for 997 days from 01-JAN-2016 to 03-FEB-2019.  The model is evaluated for 128 days from 04-FEB-2019 to 02-JUL-2019 and compared against the ground truth values.

% \added{Since market places are closed on Sundays, they are not included in our experimental setup} %\footnote{\tiny We ignore sundays in our experimentation since marketplaces are closed.}.
%first 962 days (from 2016-01-01 to 2019-02-03) as a part of the training set and for the remaining 128 days (from 2019-02-04 to 2019-08-05) we predict the crop prices and compare against the ground truths.\footnote{\tiny We ignore sundays in our experimentation since marketplaces are closed.}
%To evaluate the robustness of the crop price forecasting algorithms, we first build the model using the first 962 days (from 2016-01-01 to 2019-02-03) as a part of the training set and for the remaining 128 days (from 2019-02-04 to 2019-08-05) we predict the crop prices and compare against the ground truths.\footnote{\tiny We ignore sundays in our experimentation since marketplaces are closed.}
% First, we show results to demonstrate the importance of a multivariate model that uses marketplace arrival data,  weather parameters along with quality related features and compare against other methods that have been considered in the past.  Then, we evaluate our context-based model selection strategies. We show extensive experimentation results on the price prediction of Tomato and draw conclusions. Additionally, to demonstrate the generalization capabilities of the proposed approach, we show results on Maize crop which follow different properties when compared to Tomato.
\begin{table}
	\centering
	\caption{Distribution of Agmarknet datasets. }
	\begin{tabular}{|l||c|c|c|c|}
		\hline
		 \bf Crops & \bf Marketplaces & \bf Total Days & \bf Train (Days) & \bf Test (Days)   \\
		\hline\hline
		\bf Tomato & 7 & 1125 & 997 & 128   \\
		\hline
		\bf Maize  & 7 & 1125 & 997  & 128    \\\hline
	\end{tabular}%
	\label{tab:dataset_distribution}
    % \vspace{-10pt}
\end{table}%
% \end{comment}

\subsection{Experimental Setup}
To show the effectiveness of proposed features and context-based model selection strategies, we conduct experiments for multiple marketplaces for Tomato. For experimentation, we consider multiple factors such as \textit{model characteristics }- univariate (ARIMA, SARIMA, and Prophet) vs multivariate (MLP), \textit{features used} - only crop price vs additional features, \textit{model selection for retraining} - continuous vs context-based. Following is the list of experiments conducted:
\begin{itemize}
    \item \textbf{Baseline Models}: To establish baselines, we consider widely used univariate models such as ARIMA \cite{yunus2015arima}, SARIMA \cite{vagropoulos2016comparison}, and FB-PROPHET \cite{taylor2018forecasting}. These models only take the crop price as input to make the forecast.
    \item \textbf{Importance of Proposed Features}: To understand the effect of additional features such as AG, WD, DQ we consider multivariate models. We use feature concatenation method proposed in Sec \ref{feat_rep}(1) to prepare the feature and use MLP to conduct the experiments using continuous model retraining strategy.
    \item \textbf{Importance of Context-based Model Selection}: To show the effectiveness of context-based model selection over continuous model retraining, we conduct multiple experiments. We evaluate different context-based model selection strategies as mentioned in Sec. \ref{context_mod_sel} such as model selection based on data quality ($M^{d}$), model stability ($M^{s}$), and trend-based model selections either at marketplace level ($T^{m}$) or at crop level ($T^{c}$). In these experiments, we use two different types of feature representations namely, feature concatenation and time-series dependent feature representation as mentioned in Sec \ref{feat_rep}. 
\end{itemize}

\begin{comment}
\begin{itemize}
    \item ARIMA, SARIMA, and Prophet as baseline models which take as input only the crop price
    \item MLP model which takes as input AG and WD data along with crop prices as input
    \item MLP model which takes as input AG, WD and DQ data along with crop prices as input
\end{itemize}

In the second phase, we validate the effectiveness of the proposed context-based model retraining and conduct the following experiments
\begin{itemize}
    \item MLP with a continuous retraining
    \item MLP with a data quality based criteria for model retrieval for retraining
    \item MLP with model stability based criteria for model retrieval for retraining
    \item MLP with trend-based criteria for model retrieval for retraining
\end{itemize}
\end{comment}

To show the generalization capability of the proposed method, we also perform similar experiments for multiple marketplaces for Maize. The metrics used to evaluate the model performance and model robustness are discussed in the following section.

\subsection {Evaluation Metrics} \label{Evaluation metrics}
The performance of forecasting models is measured using Root Mean Square Error (RMSE) and Mean Absolute Percentage Error (MAPE). In addition to measuring forecasting model performance, measuring the robustness and the reliability of the deployed system over time is also a very important criteria. 
% In this work, we measure the robustness and the reliability of the model by measuring the evaluation metrics such as RMSE and MAPE over a period of time. 
% This will also help to guarantee the error bound over a longer period of time.  

As mentioned in Table \ref{tab:dataset_distribution}, we initially use 997 days data for training the model and everyday we forecast 30 days ahead crop prices. For each following day, we add the new data point and follow the training and forecasting procedure. Evaluation is done over a period of 128 days and the metrics such as average RMSE and average MAPE are computed to understand the model performance over longer period of time. The average RMSE (AR) and average MAPE (AM) are computed over $p$ days as:
% \noindent\begin{minipage}{.5\linewidth}
% \begin{equation*}\label{rmse_mod}
% \small
% AR=\frac{1}{p}\sum_{j=1}^{p}{\sqrt{\frac{1}{h}\sum_{i=1}^{h}(\hat{y}_i^{(j)} -y_i^{(j)})^2}}
% \end{equation*}
% \end{minipage}%
% \hspace{2mm}
% \begin{minipage}{.5\linewidth}
% \begin{equation*}\label{mape_mod}
% \small
% AM=\frac{100\%}{p}\sum_{j=1}^{p}{\frac{1}{h}\sum_{i=1}^{h}\left |\frac{y_i^{(j)}- \hat y_i^{(j)}}{y_i^{(j)}}\right|}
% \end{equation*}
% \end{minipage}
\begin{equation}\label{rmse_mod}
AR=\frac{1}{p}\sum_{j=1}^{p}{\sqrt{\frac{1}{h}\sum_{i=1}^{h}(\hat{y}_i^{(j)} -y_i^{(j)})^2}}
\end{equation}
\begin{equation}\label{mape_mod}
AM=\frac{100\%}{p}\sum_{j=1}^{p}{\frac{1}{h}\sum_{i=1}^{h}\left |\frac{y_i^{(j)}- \hat y_i^{(j)}}{y_i^{(j)}}\right|}
\end{equation}
where $\hat{y}_i^{(j)}$ and $y_i^{(j)}$ represent the prediction and the ground truth for $j^{th}$ test data point of $i^{th}$ day respectively. The values of $h$ and $p$ are used as $30$ and $128$ respectively in our experiments.  \added{While computing the evaluation metrics, we ignore the forecasted values when there is no ground truth data available because of missing value.}

To compare the reliability and robustness of different models graphically, we show the error variations and cumulative error distribution (CED) curves along with the average RMSE (AR) and average MAPE (AM) values for the different models. The error variation graph shows the variation of error for the forecasts, where  high fluctuations in the graph including high peaks suggest that a model is not reliable. CED curve of error values help in comparing the percentage of test data points whose \added{prediction} error falls within a particular error threshold value. A \added{higher} percentage value suggests that the model is more robust. 

\subsection{Importance of Agmarknet, Weather and Data Quality related Features}

Most of the existing techniques such as ARIMA, SARIMA, and Prophet use only historical time-series data. However, crop prices are highly dependent on the local weather conditions \added{and} market arrival \added{in addition to the} historical crop prices. Our multivariate feature representation captures marketplace specific characteristics such as weather parameters (WD) which capture the variation in the avg. temperature, avg. humidity, and the total rainfall on a daily level, along with historical crop prices and historical market place crop arrival data (AG). We compare our feature \added{concatenation based} representation (Sec. \ref{feat_rep}(1)) with existing techniques which only use the historical crop prices.

The comparative average RMSE and average MAPE evaluation metrics of the models mentioned above are shown in Table \ref{uni_vs_multi} for several marketplaces for Tomato crop.  It can be observed that the multivariate models have a lower average error for all the marketplaces when compared to the baseline models. It can also be  observed that the addition of data quality related features has improved the performance of the model as compared to using only the Agmarknet and weather-based feature representation.

\begin{figure*}[tb]
  \centering
  \begin{minipage}[b]{0.49\textwidth}
    \includegraphics[width=\textwidth]{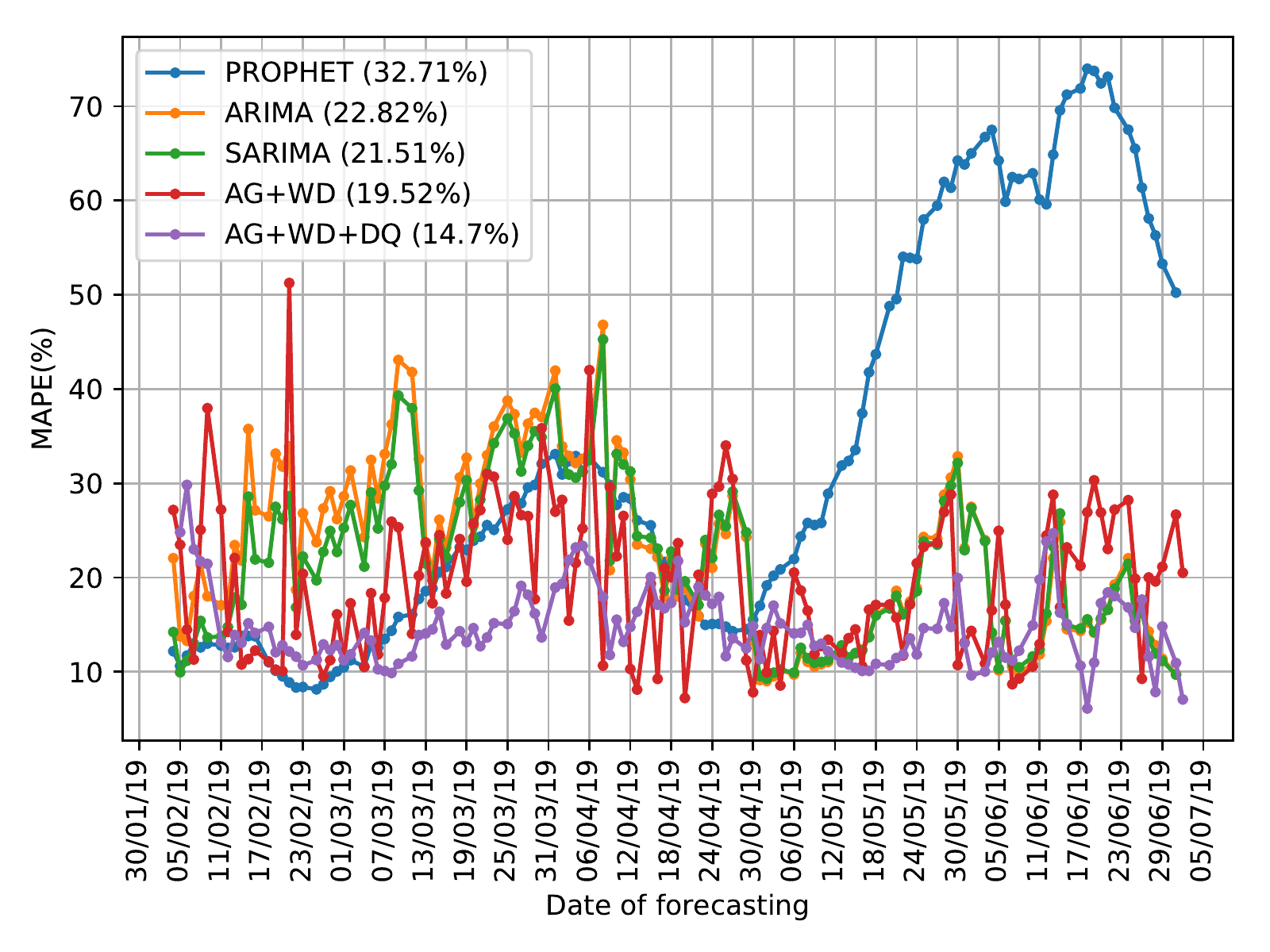}
    \vspace{-6mm}
    \caption*{(a) Error variation for daily forecast}
    % \caption{Error variation for Mulabagilu(Stage1)}
  \end{minipage}
  \hfill
  \begin{minipage}[b]{0.49\textwidth}
    \includegraphics[width=\textwidth]{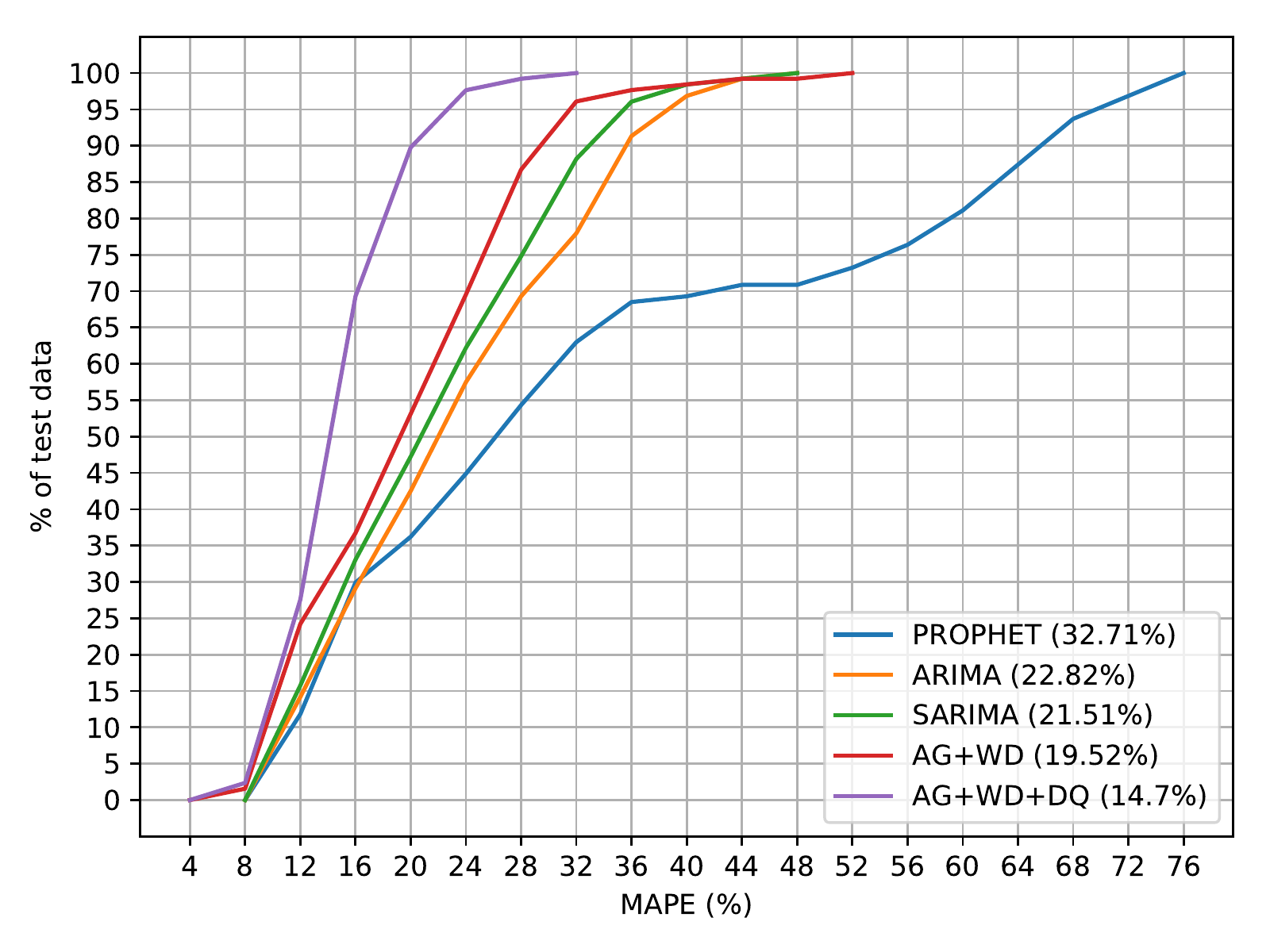}
    \vspace{-6mm}
    \caption*{(b) Cumulative error distribution curve}
    % \caption{Pareto chart for Mulabagilu(Stage1)}
  \end{minipage}
   %vspace{-2mm}
  %\caption{Error variation(a) and Cumulative error variation curve (b) for Kolar (Univariate vs. Multivariate)}
  \caption{Comparative evaluation metrics for different feature representations including univariate and multivariate models for Kolar marketplace for Tomato.}
  \vspace{-4mm}
  \label{fig:mula_2}
\end{figure*}

To qualitatively analyse the effectiveness of the proposed features, we graphically represent the error variation and CED curves in Fig.\ref{fig:mula_2}. From Fig. \ref{fig:mula_2}(a), it can be observed that for most of the forecasts, AG+WD and AG+WD+DQ models have lower errors when compared to baseline models indicating that models build using proposed features are more reliable. From the CED curve in Fig. \ref{fig:mula_2}(b), it can be observed that the AG+WD model is robust when compared to baseline models since the forecast errors fall in lower MAPE range. Further, it can also be observed that AG+WD+DQ model is better than the AG+WD model indicating that considering data quality information is helpful.
% Additionally, to analyze the importance of time-series data quality related information, we add the DQ features as mentioned in Sec. \ref{insight}, along with the Agmarknet data (AG) and weather features (WD) to enrich our feature representation. Figs. \ref{fig:mula_2}(a) and \ref{fig:mula_2}(b) show that the multivariate model (AG+WD+DQ) further improves the reliability and robustness when compared to (AG+WD) based feature representation.  
% We use a multilayer perceptron (MLP), using a continuous model retraining strategy using feature concatenation technique. %The multivariate model and continuous training method were used to demonstrate the importance of data quality related features. 

\added{In the current setup, models are continiously retrained over time. Next, we evaluate various context-based model selection strategies using two different feature representations such as feature concatenation and time-series dependent feature representation.}

\begin{table*}[tb]
\caption{Comparison between univariate and multivariate models for Tomato. AR: Average RMSE, AM: Average MAPE, AG: Agmarknet data, WD: Weather data, DQ: Data quality-based features.}
\label{uni_vs_multi}
\centering
\begin{adjustbox}{max width=\textwidth}
\begin{tabular}{|c|c|c|c|c|c|c||c|c||c|c|}
\hline
\multirow{2}{*}{\textbf{\begin{tabular}[c]{@{}c@{}}Marketplace\\  Name\end{tabular}}} & \multicolumn{2}{c|}{\textbf{Prophet\cite{taylor2018forecasting}}} & \multicolumn{2}{c|}{\textbf{ARIMA \cite{yunus2015arima}}} & \multicolumn{2}{c|}{\textbf{SARIMA \cite{vagropoulos2016comparison}}} & \multicolumn{2}{c|}{\textbf{AG+WD (MLP)}} & \multicolumn{2}{c|}{\textbf{AG+WD+DQ (MLP)}} \\ \cline{2-11} 
                                     & \textbf{AR}     & \textbf{AM}     & \textbf{AR}    & \textbf{AM}    & \textbf{AR}     & \textbf{AM}    & \textbf{AR}    & \textbf{AM}    & \textbf{AR}     & \textbf{AM}      \\ \hline \hline
% \textit{Gowribidanoor}               & 357.03            & 48.37             & 203.74           & 20.85            & 193.27            & 19.68            & 162.43           & 17.15            & \textbf{118.96}            & \textbf{12.94}     \\ \hline
\textit{Kolar}                       & 598.48            & 32.71             & 404.28           & 22.82            & 389.42            & 21.51            & 350.51           & 19.52            & \textbf{276.62}            & \textbf{14.7}      \\ \hline
\textit{Mulabagilu}                  & 568.11            & 31.37             & 413.82           & 20.87            & 417.35            & 21.48            & 361.87           & 16.79            & \textbf{292.4}             & \textbf{15.39}     \\ \hline
\textit{Chikballapur}                & 690.06            & 45.56             & 350.43           & 20.7             & 350.43            & 20.7             & 310.19           & 18.08            & \textbf{277.79}            & \textbf{15.02}     \\ \hline
% \textit{Malur}                       & 633.8             & 29.66             & 523.14           & 22.42            & 523.14            & 22.42            & 427.28           & 18.82            & \textbf{336.65}            & \textbf{16.67}     \\ \hline
\textit{Srinivasapur}                & 788.45            & 45.23             & 528.1            & 25.57            & 534.67            & 25.68            & 520.73           & 23.35            & \textbf{318.57}            & \textbf{16.16}     \\ \hline
\textit{Chintamani}                  & 716.98            & 34.71             & 539.08           & 31.0             & 550.6             & 32.39            & 602.29           & 39.45            & \textbf{462.87}            & \textbf{29.53}     \\ \hline
\textit{Bangarpet}                   & 575.39            & 31.65             & 468.56           & 26.97            & 460.37            & 26.7             & 382.55           & \textbf{20.15}            & \textbf{323.17}            & \textbf{20.15}     \\ \hline
\textit{Mysore}          & 895.51            & 42.34             & 652.3            & 26.91            & 661.1             & 27.28            & 666.5            & 25.23            & \textbf{398.69}            & \textbf{17.23}     \\ \hline
\end{tabular}
\end{adjustbox}
\vspace{-4mm}
\end{table*}

%\subsection{Importance of Context-based Model Selection using Trend Analysis}
\subsection{Importance of Context-based Model Selection}
\label{exp:context_based_model_selection}
% We evaluate different context-based model selection strategies as mentioned in Section \ref{context_mod_sel} such as model selection based on presence of outlier ($M^{d}$), model stability based model selection ($M^{s}$), and trend-based model selections either at marketplace level ($T^{m}$) or at crop level ($T^{c}$).  
Continuous retraining strategy may get adversely affected due to the presence of data quality issues such as outliers in the crop prices.  To deal with this, in data quality based model selection $M^{d}$, we dynamically retrieve the best performing model with respect to Avg RMSE from the model catalog when there is an outlier present in the recent data. Due to high fluctuations in the factors affecting crop prices, updating model daily can impact the model stability adversely. To address this issue, we enable stability based model selection ($M^{s}$), where best performing model (described in Sec. \ref{context_mod_sel}(2)) is retrieved from the catalog daily. The catalog is updated when the new model outperforms previous best performing model. Trend-based model selection happens at marketplace level $T^m$ and crop level $T^c$. In $T^{m}$, we build separate $T^{pos}$ and $T^{neg}$ models for each marketplace whereas in $T^{c}$ we just build one set of $T^{pos}$ and $T^{neg}$ models on the accumulated data of all the marketplaces and forecast. In our experiments, for trend-based model selection we use 7-day price window in Eq. \ref{eq1}. 
% Data for training the models is obtained as follows: Each data point is categorized as having a positive trend or a negative trend based on the trend score of 7-day price window as given in Eq. \ref{eq1}. All the data points categorized as having a positive trend are used to train $T^{pos}$ and vice versa. We also trained a LSTM models %\footnote{\tiny{The size of the LSTM hidden units is set to 100.}} 
% along with MLP for $T^{c}$, using the feature representation elaborated in Sec. \ref{feat_rep}.  The feature set remains the same as used in the previous subsection (AG+WD+DQ) while comparing all the model selection strategies. 
\begin{table*}[tb]
\caption{Result for context-based model selection strategies for Tomato.
AR: Average RMSE, AM: Average MAPE, $M^{d}$: data quality metric based model selection, $M^{s}$: Model stability based model selection, $T^{c}$: Trend-based modeling at crop level, $T^{m}$:  Trend-based modeling at marketplace level.}
\label{tab:context_model_sel}
\centering
\begin{adjustbox}{max width=\textwidth}
\begin{tabular}{|c||c|c||c|c||c|c||c|c||c|c|}
\hline
\multirow{2}{*}{\textbf{\begin{tabular}[c]{@{}c@{}}Marketplace\\  Name\end{tabular}}} & \multicolumn{2}{c||}{\textbf{$M^d$ (MLP)}} & \multicolumn{2}{c||}{\textbf{$M^s$ (MLP)}} & \multicolumn{2}{c||}{\textbf{$T^m$ (MLP)}} & \multicolumn{2}{c||}{\textbf{$T^c$ (MLP)}} & \multicolumn{2}{c|}{\textbf{$T^c$ (LSTM)}} \\ \cline{2-11} 
                                     & \textbf{AR}      & \textbf{AM}     & \textbf{AR}      & \textbf{AM}     & \textbf{AR}        & \textbf{AM}       & \textbf{AR}          & \textbf{AM}         & \textbf{AR}           & \textbf{AM}         \\ \hline \hline
% \textit{Gowribidanoor}               & 118.97           & 12.94           & \textbf{117.73}  & \textbf{12.97}  & 124.5              & 13.5              & \textbf{120.89}      & \textbf{12.3}       & 130.66                & 14.56               \\ \hline
\textit{Kolar}                       & 276.62           & 14.7            & 279.78           & 14.82           & 273.07             & 14.74             & \textbf{267.75}      & \textbf{14.23}      & 271.99                & 15.22               \\ \hline
\textit{Mulabagilu}                  & 291.93           & 15.36           & \textbf{286.06}  & 15.69  & 312.98             & 16.88             & 289.52      & \textbf{14.39}      & 302.82                & 15.74               \\ \hline
\textit{Chikballapur}                & 277.79           & 15.02           & 306.77           & 17.44           & 287.16             & 16.28             & 262.75      & \textbf{14.46}      & \textbf{246.77}       & 14.65      \\ \hline
% \textit{Malur}                       & 336.72           & 16.67           & \textbf{333.79}  & \textbf{15.78}  & 367.99             & 16.7              & 361.39               & 17.53               & 451.51                & 23.15               \\ \hline
\textit{Srinivasapur}                & \textbf{318.49}  & \textbf{16.15}  & 336.37           & 17.97           & 346.76             & 17.93             & 323.77               & 16.36               & 394.43                & 18.56               \\ \hline
\textit{Chintamani}                  & 462.85           & 29.54           & 446.04           & 27.27           & 438.22             & 28.64             & \textbf{399.14}      & \textbf{24.61}      & 443.52                & 25.08               \\ \hline
\textit{Bangarpet}                   & 323.3            & 20.16           & 330.22           & 20.73           & 279.19             & 17.71             & 235.63               & 13.43               & \textbf{199.06}       & \textbf{12.54}      \\ \hline
\textit{Mysore}                      & 398.85           & 17.25           & 397.65           & 16.14           & 437.82             & 17.85             & \textbf{384.41}      & \textbf{15.9}       & 461.47                & 18.25               \\ \hline
\end{tabular}
\end{adjustbox}
\vspace{-2mm}
\end{table*}

\begin{figure*}[h]
\small
  \centering
  \begin{minipage}[b]{0.49\textwidth}
    \includegraphics[width=\textwidth]{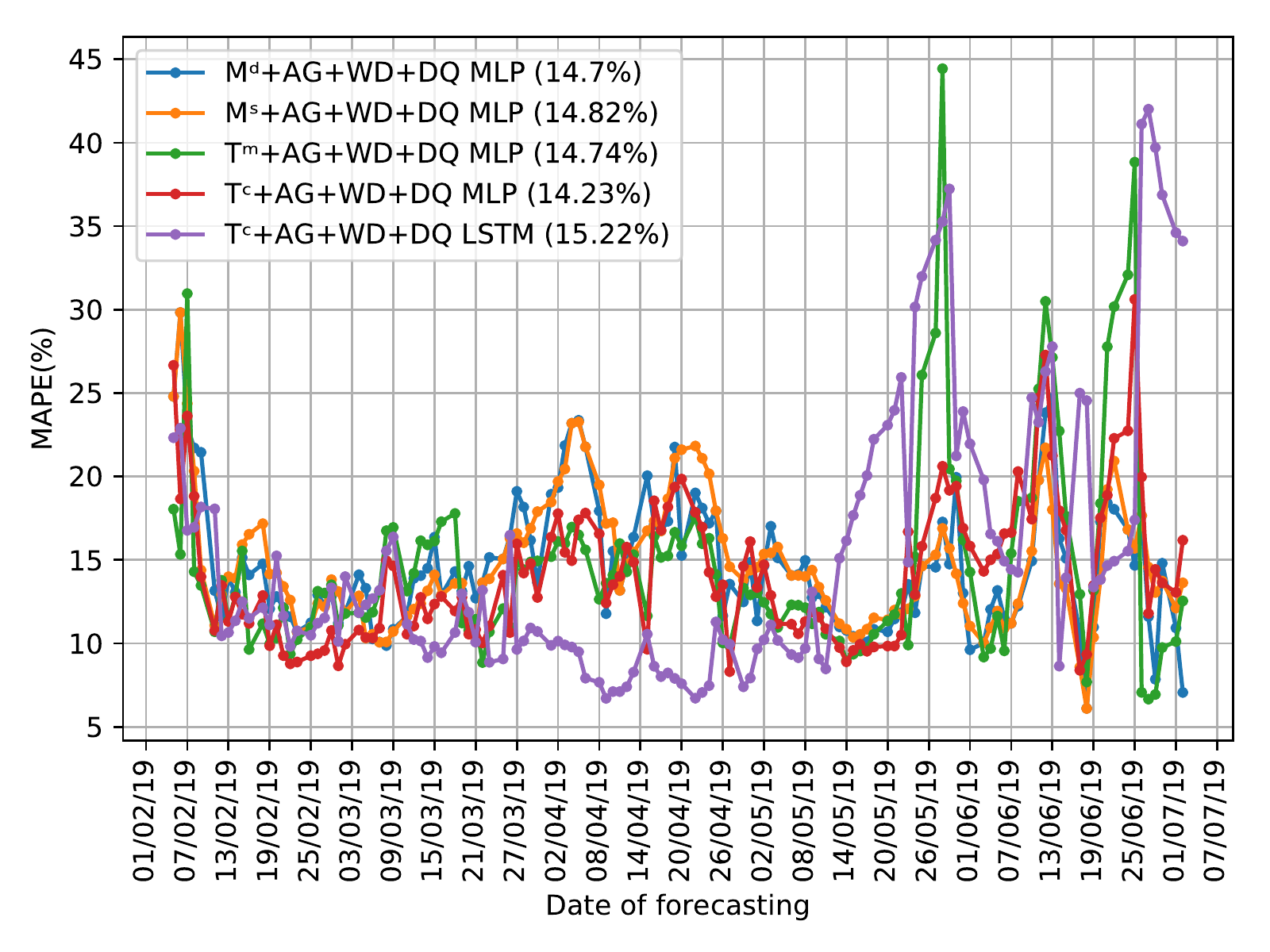}
    \vspace{-6mm}
    \caption*{(a) Error variation for daily forecast}
    % \caption{Error variation for Mulabagilu(Stage1)}
  \end{minipage}
  \hfill
  \begin{minipage}[b]{0.49\textwidth}
    \includegraphics[width=\textwidth]{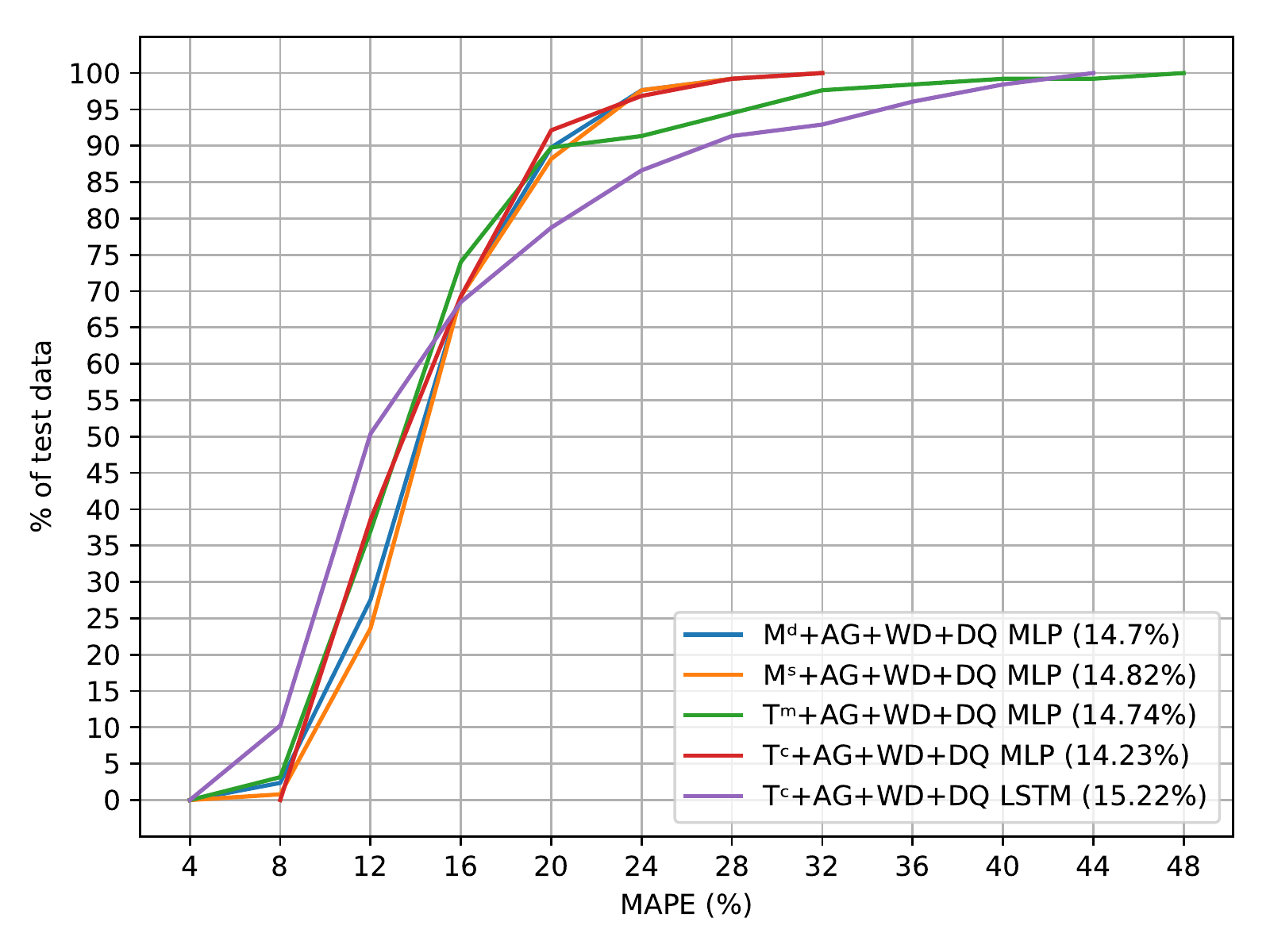}
    \vspace{-6mm}
    \caption*{(b) Cumulative error distribution curve}
    % \caption{Pareto chart for Mulabagilu(Stage1)}
  \end{minipage}
  %vspace{-2mm}
  %\caption{Error variation(a) and Cumulative error distribution curve for Kolar(Context-based model selection)}
   \caption{Comparative evaluation metrics for different context-based model selection strategies for Kolar marketplace for Tomato.}
  \label{fig:mula_3}
\end{figure*}

%We perform context-based model selection and retraining as explained in \ref{trend_analysis} and compare it against the continuous retraining evaluation strategy explained in \ref{continuous_retraining}.
Tab. \ref{tab:context_model_sel} compares the evaluation metrics for different context-based model selection strategies using the best performing AG+WD+DQ model. It can be observed that context-based model selection strategies provide robust and reliable price forecasting when compared to Tab. \ref{uni_vs_multi}. From Table \ref{tab:context_model_sel}, it can be observed that $T^{c}$ performs better than $T^{m}$ indicating that the model built on crop level data can better capture the variation in prices across marketplaces and provide better forecasting. We also evaluate the performance of crop level trend-based model selection strategy using time-series dependent feature representation ($T^c$ (LSTM)) and observe that though it performs better than $M^d$ and $M^s$, it is unable to outperform $T^c$ (MLP).

\begin{figure*}[tb]
\centering
\begin{minipage}[b]{0.49\textwidth}
    \includegraphics[width=\linewidth]{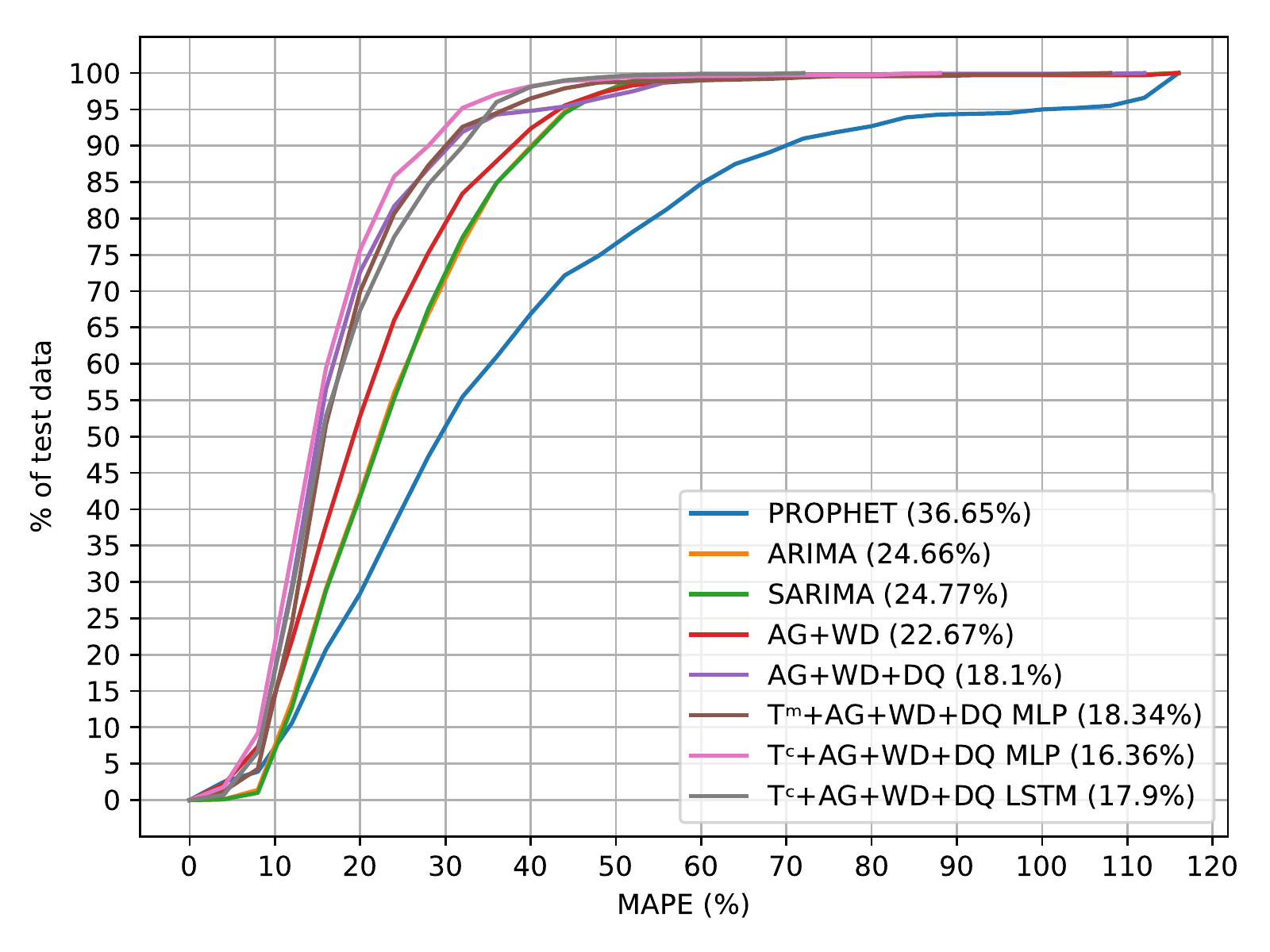}
    \vspace{-7mm}
	\caption*{(a) Tomato}
  \end{minipage}
\begin{minipage}[b]{0.49\textwidth}
	\includegraphics[width=\linewidth]{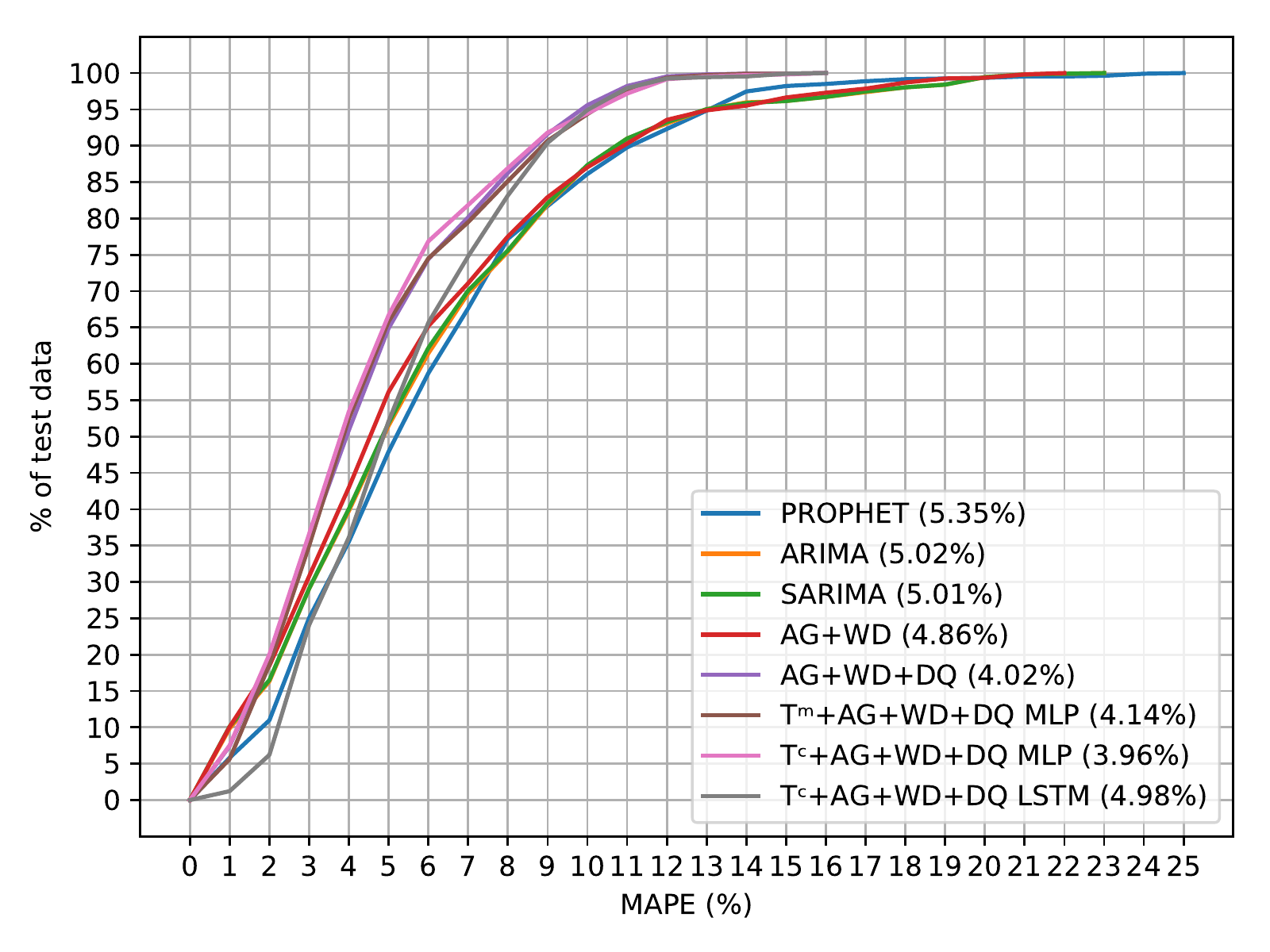}
	\vspace{-7mm}
	\caption*{(b) Maize}
  \end{minipage}
  \vspace{-2mm}
 \caption{Cumulative error distribution curve for aggregate result for all marketplaces for Tomato and Maize.}
 \label{fig:tomato_maize_all_mandi}
\end{figure*}

Fig. \ref{fig:mula_3} illustrates the detailed comparative error analysis of different strategies for Kolar marketplace. We can see from Fig. \ref{fig:mula_3}(a) that $M^s$ and $M^d$ have a relatively uniform error variations, while $T^c$ is observed to provide relatively lower MAPE\% for most of the forecasts. Fig. \ref{fig:tomato_maize_all_mandi}(a) gives the CED curve for the aggregated errors of all the marketplaces for Tomato. It can be observed that context-based model selection techniques using trend analysis ($T^{m}$ (MLP), $T^{c}$ (MLP), $T^{c}$ (LSTM)) have a lower AOC (Area over curve) indicating a lower expected error\cite{bi2003regression}. Similarly, multivariate model settings (AG+WD+DQ, AG+WD) have lower AOC when compared to baseline models. 

\added{
%Techniques such as ARIMA, SARIMA, and Prophet require the entire time-series data to perform auto-regression or seasonal decomposition to build a regressor which is similar to a continuous retraining strategy.  These techniques do not allow model management or model versioning facilities since they require complete sequential data. However, our proposed context-based model selection strategies such as model stability, data quality, and trend-based can be used when regressors do not need entire time-series data for model building. 

Techniques such as ARIMA, SARIMA, and Prophet require the entire time-series data to perform auto-regression or seasonal decomposition to build a regressor. These models have to be fit for every new data point to make new forecasts which is the same as continuous retraining strategy, hence they do not allow model management. 
% Therefore, these models can not use the other context-based model selection strategies viz. model stability, data quality, and trend-based.

%With the new data obtained because of recent crop arrivals, these models have to be fit again in order to make new forecasts, which is the same as continuous retraining strategy. These techniques require complete sequential data, hence they do not allow model management or model versioning facilities . Therefore they cannot use the other context based model selection strategies viz. model stability, data quality, and trend-based.
%which are required to enable context-based model selection strategies such as model stability, data quality and trend. 
%Therefore contexts based on model stability, data quality and trend, which require a catalog of models, cannot be used for these techniques.  
}
% \usepackage{multirow}
\begin{table*}
\caption{Comparative results of our proposed approach for Maize. 
%AR: Average RMSE, AM: Average MAPE, AG: Agmarknet data, WD: Weather data, DQ: Data quality-based features, $T^{c}$: Trend-based modeling at crop level, $T^{m}$:  Trend-based modeling at marketplace level.
}
\label{res_maize}
\begin{adjustbox}{max width=\textwidth}
\begin{tabular}{|c||c|c|c|c|c|c||c|c||c|c||c|c||c|c|}
\hline
\multirow{2}{*}{\textbf{\begin{tabular}[c]{@{}c@{}}Marketplace\\  Name\end{tabular}}} & \multicolumn{2}{c|}{\textbf{Prophet\cite{taylor2018forecasting}}} & \multicolumn{2}{c|}{\textbf{ARIMA \cite{yunus2015arima}}} & \multicolumn{2}{c|}{\textbf{SARIMA \cite{vagropoulos2016comparison}}} & \multicolumn{2}{c|}{\textbf{AG+WD+DQ (MLP)}} & \multicolumn{2}{c|}{\textbf{$T\textsuperscript{m}$ (MLP)}} & \multicolumn{2}{c|}{\textbf{$T\textsuperscript{c}$ (MLP)}} & \multicolumn{2}{c|}{\textbf{$T\textsuperscript{c}$ (LSTM)}} \\ \cline{2-15} 
                                     & \textbf{AR}       & \textbf{AM}       & \textbf{AR}      & \textbf{AM}      & \textbf{AR}       & \textbf{AM}      & \textbf{AR}        & \textbf{AM}       & \textbf{AR}           & \textbf{AM}         & \textbf{AR}           & \textbf{AM}        & \textbf{AR}           & \textbf{AM}         \\ \hline \hline
% \textit{Channagiri}                  & 67.9               & 3.47             & 43.05            & 2.27             & 40.26             & 2.11             & 58.13              & 3.05              & 38.87                 & 2.03                & \textbf{36.55}        & \textbf{1.91}      & 64.6                  & 3.39                \\ \hline
\textit{Davangere}                   & 201.34             & 8.93             & 184.23           & 8.02             & 184.23            & 8.02             & 175.49             & 7.66              & \textbf{135.86}       & \textbf{5.72}       & 140.08                & 5.95               & 143.33                & 6.23                \\ \hline
\textit{Hanagal}                     & 218.08             & 9.12             & 237.74           & 9.02             & 239.14            & 9.04             & 242.12             & 9.23              & 215.88                & 8.27                & 205.59                & 7.88               & \textbf{203.66}       & \textbf{7.72}       \\ \hline
\textit{Harappanahalli}              & 147.56             & 5.92             & 142.3            & 5.7              & 142.1             & 5.7              & 133.87             & 5.4               & 109.74                & 4.35                & \textbf{103.65}       & \textbf{4.12}      & 124.76                & 5.06                \\ \hline
\textit{Harihara}                    & \textbf{85.31}     & \textbf{3.87}    & 114.09           & 5.37             & 113.73            & 5.33             & 113.79             & 5.11              & 91.01                 & 4.21                & 95.33                 & 4.5                & 121.53                & 5.59                \\ \hline
\textit{Haveri}                      & 95.41              & 3.51             & 112.03           & 4.11             & 112.65            & 4.15             & 101.85             & 3.84              & 90.85                 & 3.28                & \textbf{79.3}         & \textbf{2.78}      & 102.97                & 3.78                \\ \hline
\textit{Honnali}                     & 96.03              & 4.03             & 80.74            & 3.33             & 80.74             & 3.33             & 83.17              & 3.48              & 80.56                 & 3.34                & \textbf{74.14}        & \textbf{3.03}      & 99.04                 & 4.16                \\ \hline
% \textit{Savanur}                     & 48.57              & 1.76             & 45.56            & 1.62             & 45.56             & 1.62             & \textbf{34.02}     & \textbf{1.36}     & 53.83                 & 2.08                & \textbf{35.3}         & \textbf{1.33}      & 85.49                 & 3.41                \\ \hline
\textit{Shiggauv}                    & 149.22             & 7.32             & 113.95           & 5.76             & 113.95            & 5.76             & 96.73              & 4.62              & 90.31        & \textbf{3.93}       & \textbf{87.44}        & 4.11      & 126.41                & 5.49                \\ \hline
\end{tabular}
\end{adjustbox}
\vspace{-3mm}
\end{table*}
% \begin{figure}[!tb]
%   \centering
%   \begin{minipage}[b]{0.49\textwidth}
%   % \vspace{-2mm}
%     \includegraphics[width=\textwidth]{figures/stage3/Error_Tomato_MAPE_Mysore(palya).pdf}
%     % \vspace{-8mm}
%     \caption*{(a)}
%   \end{minipage}
%   \hfill
%   \begin{minipage}[b]{0.49\textwidth}
%     \includegraphics[width=\textwidth]{figures/stage3/Pareto_Tomato_MAPE_Mysore(bandipalya).pdf}
%     % \vspace{-8mm}
%     \caption*{(b)}
%   \end{minipage}
%   % \vspace{-4mm}
%   \caption{Error variation(a) and pareto chart(b) for Mysore(Stage 3)}
%   \label{fig:mys_3}
% \end{figure}
\subsection{Generalization}
To show the generalization capabilities of the proposed approach, we performed crop price prediction for Maize as well. From Sec. \ref{stat_ana}, it can be noted that the residual component in Maize has a lower deviation than Tomato and this can be attributed to the fact that Maize is a cereal crop with a more regulated market than Tomato, which is primarily a cash crop. Even though the baseline models perform reasonably well, we show that proposed approach is helpful even in such scenarios where scope of improvement is less. From Tab. \ref{res_maize}, it can be observed that except for Harihara, trend-based model selection strategies have the lowest error followed by AG+WD+DQ model with continuous retraining.
% , while for the \added{Harihara} marketplace, Prophet is performing the best. 
Further, CED curves for aggregated error of all the marketplaces for Maize shown in Fig. \ref{fig:tomato_maize_all_mandi}(b) also suggests that the trend-based model selection strategy results in lower forecasting errors.
%  \subsubsection{Results for Maize}
%  \begin{figure*}[ht!]
%             \includegraphics[width=.3\textwidth]{figures/Histogram_Tomato_MAPE_MLP_False_False_True_False_False.png}\hfill
%             \includegraphics[width=.3\textwidth]{figures/Histogram_Tomato_MAPE_MLP_True_False_True_False_False.png}\hfill
%             \includegraphics[width=.3\textwidth]{figures/Histogram_Tomato_MAPE_MLP_True_False_True_False_True.png}
%             \caption{Normal settings}
%             \caption{Continuous training}
%             \caption{conditional continuous training}
%         \end{figure*}

%   \subsection{Comparative Results for different feature representation and different Models}
%  \subsubsection{Ablation Study}
\section{Conclusions \& Future Work} % and Future Work}
Crop price forecasting has been a well-studied problem in the time-series analysis domain over the years. In our experimentation, it was identified that the widely used time-series forecasting approaches (ARIMA, SARIMA, and Prophet) are affected significantly by the data quality issues, which is a predominant factor in emerging economies such as India. To address these shortcomings, we introduced a novel feature representation that captures the weather condition, historical marketplace arrival, and data quality information for robust crop price forecasting. Furthermore, we have also proposed a framework for model selection based on the context for robust crop price forecasting. Experiments indicate that the trend-based model selection strategy is useful compared to existing techniques especially in the case of highly fluctuating crop prices.
%In the future we want to explore better methodologies for fixing data quality issues by utilising the information available in the nearby marketplaces. 
In the future work, we want to focus on two aspects -- efficient data quality improvement and enhanced context-based model selection. To mitigate the data quality issues efficiently, we plan to explore techniques such as leveraging information from nearby market place to impute missing values as mentioned in \cite{Ma2018AnIP}. Similarly, we plan to explore more context-based rules and usage of meta-learning techniques like model stacking to choose the best possible strategy from the proposed strategies and obtain better performance.

\bibliographystyle{IEEEtran}
\bibliography{sample}

\begin{thebibliography}{10}
\providecommand{\url}[1]{#1}
\csname url@samestyle\endcsname
\providecommand{\newblock}{\relax}
\providecommand{\bibinfo}[2]{#2}
\providecommand{\BIBentrySTDinterwordspacing}{\spaceskip=0pt\relax}
\providecommand{\BIBentryALTinterwordstretchfactor}{4}
\providecommand{\BIBentryALTinterwordspacing}{\spaceskip=\fontdimen2\font plus
\BIBentryALTinterwordstretchfactor\fontdimen3\font minus
  \fontdimen4\font\relax}
\providecommand{\BIBforeignlanguage}[2]{{%
\expandafter\ifx\csname l@#1\endcsname\relax
\typeout{** WARNING: IEEEtran.bst: No hyphenation pattern has been}%
\typeout{** loaded for the language `#1'. Using the pattern for}%
\typeout{** the default language instead.}%
\else
\language=\csname l@#1\endcsname
\fi
#2}}
\providecommand{\BIBdecl}{\relax}
\BIBdecl

\bibitem{Bao2004ForecastingSP}
Y.~Bao, Y.~Lu, and J.~Zhang, ``Forecasting stock price by svms regression,'' in
  \emph{ICOAI}, 2004.

\bibitem{Amarasinghe2017DeepNN}
K.~Amarasinghe, D.~L. Marino, and M.~Manic, ``Deep neural networks for energy
  load forecasting,'' 2017.

\bibitem{Li2017DiffusionCR}
Y.~Li, R.~Yu, C.~Shahabi, and Y.~Liu, ``Diffusion convolutional recurrent
  neural network: Data-driven traffic forecasting,'' in \emph{ICLR}, 2017.

\bibitem{Shahhosseini2020ForecastingCY}
M.~Shahhosseini, G.~Hu, and S.~V. Archontoulis, ``Forecasting corn yield with
  machine learning ensembles,'' \emph{arXiv:2001.09055}, 2020.

\bibitem{Qiu2015RobustEO}
H.~Qiu, S.~Xu, F.~Han, H.~Liu, and B.~Caffo, ``Robust estimation of transition
  matrices in high dimensional heavy-tailed vector autoregressive processes,''
  in \emph{ICML'15}, 2015.

\bibitem{Melnyk2016EstimatingSV}
I.~Melnyk and A.~Banerjee, ``Estimating structured vector autoregressive
  models,'' in \emph{ICML}, 2016.

\bibitem{KIM2003307}
K.-j. Kim, ``Financial time series forecasting using support vector machines,''
  \emph{Neurocomputing}, 2003.

\bibitem{LI2014996}
J.~Li and W.~Chen, ``Forecasting macroeconomic time series: Lasso-based
  approaches and their forecast combinations with dynamic factor models,''
  \emph{International Journal of Forecasting}, 2014.

\bibitem{rsta2011}
S.~Roberts, M.~Osborne, M.~Ebden, S.~Reece, N.~Gibson, and S.~Aigrain,
  ``Gaussian processes for time-series modelling,'' \emph{PHILOS T R SOC A},
  2013.

\bibitem{SiamiNamini2018ForecastingEA}
S.~Siami-Namini and A.~S. Namin, ``Forecasting economics and financial time
  series: Arima vs. lstm,'' \emph{arXiv preprint arXiv:1803.06386}, 2018.

\bibitem{OliveiraT14}
M.~Oliveira and L.~Torgo, ``Ensembles for time series forecasting,'' in
  \emph{ACML}, 2014.

\bibitem{CerqueiraTPS17}
V.~Cerqueira, L.~Torgo, F.~Pinto, and C.~Soares, ``Arbitrated ensemble for time
  series forecasting,'' in \emph{ECML PKDD}, 2017.

\bibitem{ShenBAW13}
W.~Shen, V.~Babushkin, Z.~Aung, and W.~L. Woon, ``An ensemble model for
  day-ahead electricity demand time series forecasting,'' in \emph{Proceedings
  of the Fourth International Conference on Future energy systems}, 2013.

\bibitem{bame_2008}
A.~Dieng, ``Alternative forecasting techniques for vegetable prices in
  senegal,'' \emph{Senegalese journal of agricultural research}, 2008.

\bibitem{Yercan2012}
M.~Yercan and H.~Adanacioglu, ``An analysis of tomato prices at wholesale level
  in turkey: An application of sarima model,'' \emph{Custos e Agronegocio},
  2012.

\bibitem{Gloria2013}
G.~Martin-Rodriguez and J.~Caceres-Hernandez, ``Canary tomato export prices:
  comparison and relationships between daily seasonal patterns,'' \emph{SJAR},
  2013.

\bibitem{Nasira2012VegetablePP}
G.~M. Nasira and N.~Hemageetha, ``Vegetable price prediction using data mining
  classification technique,'' \emph{PRIME}, 2012.

\bibitem{Hemageetha2013RadialBF}
N.~Hemageetha and G.~Nasira, ``Radial basis function model for vegetable price
  prediction,'' in \emph{PRIME}, 2013.

\bibitem{Ouyang2019}
H.~Ouyang, X.~Wei, and Q.~Wu, ``Agricultural commodity futures prices
  prediction via long-and short-term time series network,'' \emph{Journal of
  Applied Economics}, 2019.

\bibitem{Xiong2018SeasonalFO}
T.~Xiong, C.~Li, and Y.~Bao, ``Seasonal forecasting of agricultural commodity
  price using a hybrid stl and elm method: Evidence from the vegetable market
  in china,'' \emph{Neurocomputing}, 2018.

\bibitem{taylor2018forecasting}
S.~J. Taylor and B.~Letham, ``Forecasting at scale,'' \emph{The American
  Statistician}, 2018.

\bibitem{chakraborty2016predicting}
S.~Chakraborty, A.~Venkataraman, S.~Jagabathula, and L.~Subramanian,
  ``Predicting socio-economic indicators using news events,'' in \emph{KDD},
  2016.

\bibitem{Ma2018AnIP}
W.~Ma, K.~Nowocin, N.~Marathe, and G.~H. Chen, ``An interpretable produce price
  forecasting system for small and marginal farmers in india using
  collaborative filtering and adaptive nearest neighbors,'' in \emph{ICTD '19},
  2018.

\bibitem{madaan2019price}
L.~Madaan, A.~Sharma, P.~Khandelwal, S.~Goel, P.~Singla, and A.~Seth, ``Price
  forecasting \& anomaly detection for agricultural commodities in india,'' in
  \emph{COMPASS}, 2019.

\bibitem{Zhang2020ForecastingAC}
D.~Zhang, S.~Chen, L.~Liwen, and Q.~Xia, ``Forecasting agricultural commodity
  prices using model selection framework with time series features and forecast
  horizons,'' \emph{IEEE Access}, 2020.

\bibitem{Kantanantha2010}
N.~Kantanantha, N.~Serban, and P.~Griffin, ``Yield and price forecasting for
  stochastic crop decision planning,'' \emph{JABES}, 2010.

\bibitem{junninen2004methods}
H.~Junninen, H.~Niska, K.~Tuppurainen, J.~Ruuskanen, and M.~Kolehmainen,
  ``Methods for imputation of missing values in air quality data sets,''
  \emph{Atmospheric Environment}, 2004.

\bibitem{barbato2011features}
G.~Barbato, E.~Barini, G.~Genta, and R.~Levi, ``Features and performance of
  some outlier detection methods,'' \emph{Journal of Applied Statistics}, 2011.

\bibitem{schlitzer1995testing}
G.~Schlitzer, ``Testing the stationarity of economic time series: further monte
  carlo evidence,'' \emph{Ricerche Economiche}, 1995.

\bibitem{fumi2013fourier}
A.~Fumi, A.~Pepe, L.~Scarabotti, and M.~M. Schiraldi, ``Fourier analysis for
  demand forecasting in a fashion company,'' \emph{IJEBM}, 2013.

\bibitem{thomas2019time}
K.~Thomas, ``Time series prediction for stock price and opioid incident
  location,'' Ph.D. dissertation, Arizona State University, 2019.

\bibitem{yunus2015arima}
K.~Yunus, T.~Thiringer, and P.~Chen, ``Arima-based frequency-decomposed
  modeling of wind speed time series,'' \emph{IEEE Trans. on Power Syst.},
  2015.

\bibitem{vagropoulos2016comparison}
S.~I. Vagropoulos, G.~Chouliaras, E.~G. Kardakos, C.~K. Simoglou, and A.~G.
  Bakirtzis, ``Comparison of sarimax, sarima, modified sarima and ann-based
  models for short-term pv generation forecasting,'' in \emph{ENERGYCON 2016}.

\bibitem{bi2003regression}
J.~Bi and K.~P. Bennett, ``Regression error characteristic curves,'' in
  \emph{ICML-03}, 2003.

\end{thebibliography}
\end{document}